%% file: main.tex
\documentclass{article}

\usepackage[english]{babel}

\usepackage[letterpaper,top=2cm,bottom=2cm,left=3cm,right=3cm,marginparwidth=1.75cm]{geometry}

\usepackage[utf8]{inputenc}
\usepackage[T1]{fontenc}

\usepackage{amsmath}
\usepackage{amssymb}
\usepackage{graphicx}
\usepackage[colorlinks=true, allcolors=blue]{hyperref}    

\usepackage{braket}
\usepackage{qcircuit}
\usepackage[ruled,linesnumbered]{algorithm2e} 

\usepackage{pgfplots}
\pgfplotsset{compat=newest}
\pgfplotsset{scaled y ticks=false}
\usepgfplotslibrary{groupplots}
\usepgfplotslibrary{dateplot}
\usepackage{tikz}

\usepackage{arydshln}

\usepackage{caption}
\usepackage{subcaption}

\usepackage{authblk}
\usepackage{cite}

\DeclareMathOperator{\diag}{diag}
\DeclareMathOperator*{\argmax}{arg\,max}

\title{A quantum-classical reinforcement learning model\\ to play Atari games}
\author[1]{Dominik Freinberger}
\author[1]{Julian Lemmel}
\author[1]{Radu Grosu}
\author[2]{Sofiene Jerbi}
\affil[1]{Institut f\"ur Technische Informatik, Technische Universit\"at Wien, Austria}
\affil[2]{Dahlem Center for Complex Quantum Systems, Freie Universit\"at Berlin, Germany}

\begin{document}

\maketitle

\begin{abstract}
\noindent Recent advances in reinforcement learning have demonstrated the potential of quantum learning models based on parametrized quantum circuits as an alternative to deep learning models. On the one hand, these findings have shown the ultimate exponential speed-ups in learning that full-blown quantum models can offer in certain -- artificially constructed -- environments.\break On the other hand, they have demonstrated the ability of experimentally accessible PQCs to solve OpenAI Gym benchmarking tasks. However, it remains an open question whether these near-term QRL techniques can be successfully applied to more complex problems exhibiting high-dimensional observation spaces. In this work, we bridge this gap and present a hybrid model combining a PQC with classical feature encoding and post-processing layers that is capable of tackling Atari games. A classical model, subjected to architectural restrictions similar to those present in the hybrid model is constructed to serve as a reference. Our numerical investigation demonstrates that the proposed hybrid model is capable of solving the Pong environment and achieving scores comparable to the classical reference in Breakout. Furthermore, our findings shed light on important hyperparameter settings and design choices that impact the interplay of the quantum and classical components. This work contributes to the understanding of near-term quantum learning models and makes an important step towards their deployment in real-world RL scenarios.
\end{abstract}


\section{Introduction}
\label{sec:introduction}
Reinforcement learning (RL) has had a profound impact on the machine learning (ML) landscape, thanks to the integration of deep neural networks (DNNs). Deep RL led to a series of advancements in mastering complex decision-making tasks in Atari games \cite{mnih_playing_2013, mnih_human-level_2015}, Go \cite{silver_mastering_2016, silver_mastering_2017}, StarCraft II \cite{vinyals_grandmaster_2019}, and Dota 2 \cite{openai_dota_2019}, where agents have surpassed human-level performance. However, despite its capabilities, RL remains one of the most computationally demanding areas within ML, and would substantially benefit from enhancements in computational efficiency.\\

In the pursuit of more efficient RL methods, quantum computing (QC) is a potential remedy. QC is a paradigm for computing in which information is processed based on the principles of quantum mechanics \cite{Nielsen_Chuang_2010}. By harnessing quantum mechanical phenomena such as superposition and entanglement and using them as computational resources, QC tries to gain computational advantages in certain problem classes. Here, quantum machine learning (QML) based on variational quantum algorithms (VQAs) \cite{cerezo_variational_2021} is a promising application, compatible with current noisy intermediate-scale quantum (NISQ) devices \cite{preskill_quantum_2018}. VQAs optimize parametrized quantum circuits (PQCs), quantum circuits composed of quantum gates that can be tuned via classical hardware to achieve a desired computational outcome. Similarly to classical DNNs, the PQC acts as a function approximator that is trained via classical optimization techniques, in a hybrid quantum-classical manner. PQC-based QML models and hybrid approaches that leverage the strengths of both classical and quantum components have already demonstrated significant results in both supervised \cite{mitarai_quantum_2018, farhi_classification_2018, havlicek_supervised_2019, schuld_quantum_2019, schuld_circuit-centric_2020, mari_transfer_2020} and unsupervised \cite{amin_quantum_2018, lloyd_quantum_2018, chakrabarti_quantum_2019, zoufal_quantum_2019} machine learning tasks.\\

Research in quantum reinforcement learning (QRL) with PQCs has only recently started gaining traction \cite{chen_variational_2020, lockwood_reinforcement_2020, jerbi_parametrized_2021, skolik_quantum_2022, chen_quantum_2023, meyer2022survey}. Nonetheless, the early results have already made great strides in showing the potential of QRL methods based on PQCs. From a theory standpoint, they have shown the existence of RL tasks where PQC models could achieve exponential speed-ups in learning compared to all classical models, including deep neural networks \cite{jerbi_parametrized_2021, skolik_quantum_2022}. These environments remain however artificially constructed using cryptographic functions, and the PQC models used rely on large-scale, fault-tolerant implementations of Shor's algorithm. From a more practical standpoint, these works have also established that small-scale PQCs are capable of solving iconic benchmarking environments from OpenAI Gym \cite{chen_variational_2020, lockwood_reinforcement_2020, jerbi_parametrized_2021, skolik_quantum_2022}, which have been successfully implemented on current quantum hardware \cite{hsiao2022unentangled, meyer2023quantum}. Despite these advances, research in this domain remains sparse, with only a few existing works \cite{lockwood_playing_2021, chen_deep-q_2023, chen_variational_2022, chen_efficient_2024} addressing the challenges of applying QRL agents to environments with high-dimensional observations, as present in many real-world applications. This leaves as an open question whether QRL agents can effectively learn and operate in these complex settings, and, if so, which model design choices are most crucial in achieving a good learning performance, and how these models compare to their classical counterparts.

\begin{figure}
    \centering
    \includegraphics[width=1.0\textwidth]{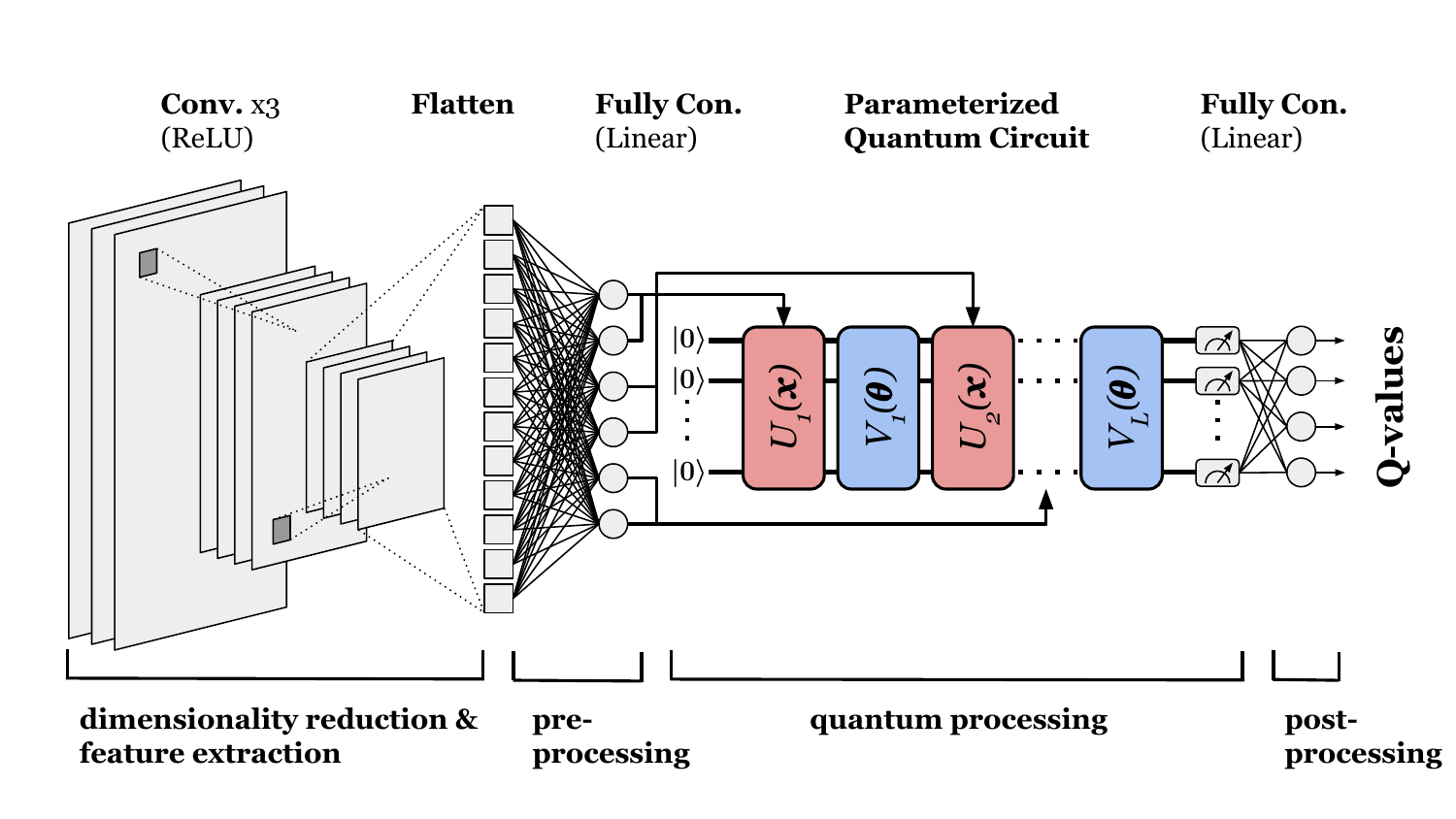}
    \caption{\textbf{The hybrid quantum-classical architecture}. Three convolutional layers reduce the high-dimensional input and extract a low number of features which are further combined and reduced by a linear pre-processing layer to create a highly informative latent representation. The PQC encodes these latent features and outputs the expectation values of local Pauli-$Z$ measurements. The output of the PQC is further post-processed by another fully connected layer with linear activation to match the Q-value magnitudes and action space dimension.}
    \label{fig:hybrid-quantum-classical-model}
\end{figure}

\subsection{Contributions}
In this work, we propose a hybrid quantum-classical model and analyze its performance in RL environments with high-dimensional observation spaces. As a testbed, we consider the Atari 2600 game environments Pong and Breakout, often used as benchmarks for deep RL \cite{machado_revisiting_2018}. Our main contribution is to show that a hybrid quantum-classical model is capable of solving the Pong environment according to its specifications \cite{noauthor_atariage_nodate} and reaching a performance similar to a classical reference in the game Breakout. This is in stark contrast with the findings of \cite{lockwood_playing_2021}, which reported that hybrid models are unable to learn in the Atari environments due to a lack of expressivity. Our second contribution is an in-depth analysis of the design choices that are responsible for an increase in performance. In particular, we find that rescaling rewards and adjusting the learning rate in the output layer can have a notable impact on learning performance, which we relate to the different nature of Q-function landscapes that hybrid and classical models will learn. This research extends previous studies that first demonstrated the learning capabilities of quantum and hybrid models in an approximate Q-learning setting \cite{chen_variational_2020, lockwood_reinforcement_2020, jerbi_parametrized_2021, chen_deep-q_2023} in simple benchmarking environments from the OpenAI Gym \cite{brockman_openai_2016}. While our results do not show any kind of advantage of quantum-enhanced models over fully classical models, they nonetheless make an important step in this direction by showcasing how an interplay of classical feature extraction and quantum processing can be successfully combined to solve complex, high-dimensional RL problems.

\subsection{Related work}
The first efforts to use PQC-based models as function approximators in RL were made in \cite{chen_variational_2020}, where quantum agents demonstrated learning capabilities in the simple benchmark environment \texttt{Frozen} \texttt{Lake} from the OpenAI Gym \cite{brockman_openai_2016} using an approximate Q-learning approach. Similarly, \cite{lockwood_reinforcement_2020} benchmarked PQCs as Q-function approximators in \texttt{CartPole} and \texttt{Blackjack}, proposing different encoding schemes and confirming that PQC models can perform comparably to DNNs. In \cite{jerbi_parametrized_2021}, PQCs were successfully used as policy function approximators to solve \texttt{CartPole}, \texttt{MountainCar}, and \texttt{Acrobot}, highlighting data re-uploading \cite{perez-salinas_data_2020, schuld_effect_2021} and trainable scaling weights on inputs and observables as key design choices. Concurrently, \cite{skolik_quantum_2022} adopted a similar architecture for Q-learning in \texttt{FrozenLake} and \texttt{CartPole}, stressing the importance of trainable weights on observables for matching the range of Q-values. In \cite{jerbi_parametrized_2021} and \cite{skolik_quantum_2022}, the authors demonstrate a theoretical quantum advantage of quantum agents over classical learners in both policy- and value-based reinforcement learning settings, building on results on supervised learning based on the discrete logarithm problem \cite{liu_rigorous_2021}, while the former also provides numerical evidence of an empirical advantage over DNNs in specifically crafted PQC-based environments.\\

Most of these studies deal with environments characterized by low-dimensional observations. However, \cite{chen_deep-q_2023} tackled a 20 × 20 dimensional maze environment, using a hybrid model with a classical DNN to pre-process inputs, demonstrating adaptability to higher-dimensional problems. Likewise, \cite{chen_variational_2022} introduced a tensor-network-based variational quantum circuit (TN-VQC) architecture for dimensionality reduction of high-dimensional inputs. A step closer to real-world problems is taken in \cite{lockwood_playing_2021} extending previous work \cite{lockwood_reinforcement_2020} by proposing a hybrid model combining a classical convolutional network and quantum circuits. The authors study the performance of the hybrid architecture in the Atari games \texttt{Pong} and \texttt{Breakout}, where an observation consists of approximately 100,000 variables. Although the hybrid model achieves non-trivial results in \texttt{CartPole}, it fails to learn in the high-dimensional Atari game environments. In our work, we draw inspiration from several of these studies, particularly those that explore hybrid quantum-classical models. We overcome the shortcomings of \cite{lockwood_playing_2021} and demonstrate that hybrid quantum-classical agents can successfully learn in \texttt{Pong} and \texttt{Breakout}.


\section{Quantum reinforcement learning}
This section gives a concise introduction to quantum computing and its application in machine learning. It further introduces the hybrid quantum-classical model used as a Q-function approximator and its implementation in the approximate Q-learning framework.

\subsection{Quantum computing}
The basic unit of quantum information is a qubit, a two-level quantum system represented by a two-dimensional complex Hilbert space $\mathcal{H} = \mathbb{C}^2$. The two orthogonal states $\ket{0} = (1,0)^T$ and $\ket{1} = (0,1)^T$ form a basis of this space and are referred to as computational basis states\footnote{We use Dirac-notation, where vectors and their conjugate transpose are denoted as $\ket{\psi}$ and $\bra{\psi}$ respectively and their inner product as $\braket{\psi|\phi}$.}. A general state of a qubit is described by a unit vector $\ket{\psi} := a \ket{0} + b \ket{1} \in \mathcal{H}$, where $a, b \in \mathbb{C}$ satisfy $|a|^2 + |b|^2 = 1$. This generalizes to $n$ qubits, where the combined Hilbert space for an $n$-qubit system is $\mathcal{H}^{\otimes n}$ and a general state is described by a superposition of all $2^n$ possible combinations of its qubits' basis states. The states of qubits are manipulated through unitary operators $U$ acting on $\mathcal{H}^{\otimes n}$, called quantum gates. An important set of single-qubit gates are the Pauli gates denoted as $X$, $Y$ and $Z$ that give rise to the Pauli rotation gates $R_x$, $R_y$ and $R_z$, which are famous examples of parameterized gates:

\begin{align}
\label{eqn:pauli-gates}
&R_x(\theta) = e^{-i \frac{\theta}{2} X},
X=\begin{bmatrix} 0 & 1 \\ 1 & 0 \end{bmatrix} & 
&R_y(\theta) = e^{-i \frac{\theta}{2} Y},
Y=\begin{bmatrix} 0 & -i \\ i & 0 \end{bmatrix} & 
&R_z(\theta) = e^{-i \frac{\theta}{2} Z},
Z=\begin{bmatrix} 1 & 0 \\ 0 & -1 \end{bmatrix},
\end{align}

with rotation angels $\theta \in \mathbb{R}$. An important $2$-qubit gate is the $CZ := \diag(1,1,1,-1)$ or Controlled-$Z$ gate. It is a controlled gate that only affects the second qubit, or target qubit, if the first qubit, the control qubit, is in the $\ket{1}$ state and is used to introduce entanglement between multiple qubits.\\

Unlike a classical bit, the state of a qubit is accessible only through quantum measurement, performed using an observable represented by a Hermitian operator $\boldsymbol{\mathcal{M}}=\boldsymbol{\mathcal{M}}^\dag$. The spectral decomposition of $\boldsymbol{\mathcal{M}} = \sum_m m P_m$ into eigenvalues $m$ and corresponding orthogonal projections $P_m$, defines the possible outcomes of the measurement. According to the Born rule, when a quantum state $\ket{\psi}$ is measured, the outcome $m$ occurs with probability $p(m) = \braket{\psi | P_m | \psi} = \braket{P_m}$ and the state is projected onto $P_m \ket{\psi} / \sqrt{p(m)}$. The expectation value of a measurement is defined as:

\begin{equation}
\label{eqn:expectation-value}
E(\boldsymbol{\mathcal{M}}) = \sum_m p(m) m =: \braket{\boldsymbol{\mathcal{M}}}.
\end{equation}

Considering the spectrum of the Pauli $Z$-matrix, the expectation value of a computational basis measurement of a single qubit, where $Z$ is the observable acting on that qubit, falls within the interval $[-1, 1]$.

\begin{figure}
  \centering
  \includegraphics[width=0.75\linewidth]{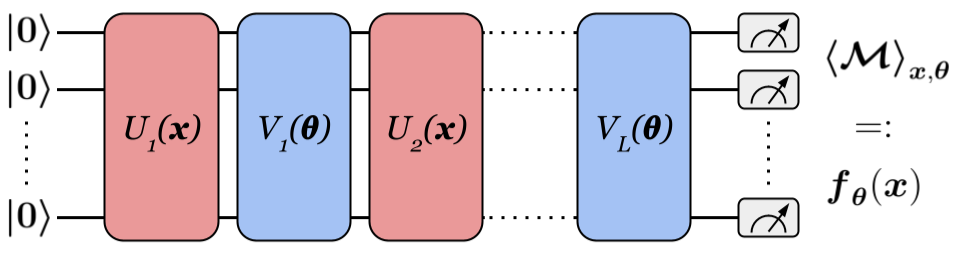}
  \caption{\textbf{A parameterized quantum circuit as a machine learning model.} A feature vector $\boldsymbol{x}$ is encoded into the quantum system in its trivial state $\ket{0}^{\otimes n}$ via the repeated encoding unitaries $U_l(\boldsymbol{x})$ (red). Intermediate variational unitaries $V_l(\boldsymbol{\theta})$ (blue) enable the training of the circuit. The output of the model $f_{\boldsymbol{\theta}}(\boldsymbol{x})$ is the expectation value $\braket{\boldsymbol{\mathcal{M}}}_{\boldsymbol{x}, \boldsymbol{\theta}}$ of a (or multiple) observables (e.g., a Pauli-$Z$ observable on each qubit) measured at the end of the circuit.}
  \label{fig:pqc}
\end{figure}

\subsection{The hybrid quantum-classical model}
\label{sec:hybrid-model}
Having introduced quantum computing we now move to its application in ML by defining the hybrid quantum-classical model as illustrated in Figure~\ref{fig:hybrid-quantum-classical-model}. Given an initial state $\ket{\psi}$ of an $n$-qubit quantum system, for example the trivial state $\ket{0}^{\otimes n}$, a PQC applies the parameter-dependent unitary transformation $U(\boldsymbol{x}, \boldsymbol{\theta})$ to its qubits, which is defined as

\begin{equation}
\label{eqn:pqc}
    U(\boldsymbol{x}, \boldsymbol{\theta}) = \prod_{l=1}^L V_l(\boldsymbol{\theta}) U_l(\boldsymbol{x}),
\end{equation}

where the $U_l(\boldsymbol{x})$ encode parts of the feature vector $\boldsymbol{x} \in R^d$ into the quantum state using $R_x$ gates. The $V_l(\boldsymbol{\theta})$ consist of $R_x$, $R_y$ and $R_z$ gates that depend on adjustable parameters $\boldsymbol{\theta} \in \mathbb{R}^k$ that are optimized during the training process by classical hardware. The repeated encoding of the feature vector $\boldsymbol{x}$, referred to as data re-uploading \cite{perez-salinas_data_2020}, is known to enhance the expressivity of PQC-based machine learning models \cite{schuld_effect_2021}. The state of the system resulting from the application of the circuit is $U(\boldsymbol{x}, \boldsymbol{\theta}) \ket{0}^{\otimes n} = \ket{\psi(\boldsymbol{x}, \boldsymbol{\theta})}$. The expectation value of some measurement observable $\boldsymbol{\boldsymbol{\mathcal{M}}}$ on the quantum system prepared by the PQC is 

\begin{equation}
\label{eqn:determ-qml-model}
    \braket{\boldsymbol{\mathcal{M}}}_{\boldsymbol{x}, \boldsymbol{\theta}} = \braket{\psi(\boldsymbol{x}, \boldsymbol{\theta}) | \boldsymbol{\mathcal{M}} | \psi(\boldsymbol{x}, \boldsymbol{\theta})} =: f_{\boldsymbol{\theta}}(\boldsymbol{x}).
\end{equation}

This defines a deterministic quantum machine learning model $f_{\boldsymbol{\theta}}(\boldsymbol{x})$, which, given input $\boldsymbol{x}$ and parameters $\boldsymbol{\theta}$, produces a deterministic value $\braket{\boldsymbol{\mathcal{M}}}_{\boldsymbol{x}, \boldsymbol{\theta}}$. While quantum models as defined in Equation~\ref{eqn:determ-qml-model} have the potential for processing various kinds of data, the direct encoding of high-dimensional inputs, such as images, into quantum circuits via single-qubit rotations is currently infeasible. The low qubit count of NISQ hardware, as well as the computational demands of simulating larger qubit systems, render it impossible to encode vast amounts of data into a quantum circuit. To address these limitations, we incorporate classical convolutional layers as trainable dimensionality reduction components. These classical layers pre-process the high-dimensional input, transforming it into fewer, highly informative features for the PQC. We define the hybrid model as:

\begin{equation}
\label{eqn:hybrid-model}
    \mathcal{Q}_{hybrid} = L_{\boldsymbol{w}_{out}}(f_{\boldsymbol{\theta}}(L_{\boldsymbol{w}_{in}}(\boldsymbol{\Tilde{x}}))),
\end{equation}

where $\mathcal{Q}_{hybrid}$ represents the entire hybrid model. Here, $L_{\boldsymbol{w}_{in}} : \mathbb{R}^{n_{in}} \rightarrow \mathbb{R}^{n_q}$ represents the classical pre-processing layers that map the raw unprocessed input data $\boldsymbol{\Tilde{x}}$ to a format suitable for the quantum circuit. $f_{\boldsymbol{\theta}} : \mathbb{R}^{n_q} \rightarrow \mathbb{R}^{n_q}$ is the core PQC as defined in Equation~\ref{eqn:determ-qml-model}, and $L_{\boldsymbol{w}_{out}} : \mathbb{R}^{n_q} \rightarrow \mathbb{R}^{n_{out}}$ is a classical post-processing layer that produces the final output. This hybrid approach leverages the strengths of both quantum and classical computing, allowing us to optimize the entire model through a combination of quantum gate parameters $\boldsymbol{\theta}$ and classical neural network weights $\boldsymbol{w}_{in}$, $\boldsymbol{w}_{out}$.

\subsection{Hybrid model architecture}
\label{sec:hybrid-model-architecture}
In this section, we elaborate on the specific architecture of the hybrid model used in our experiments. Our hybrid model design is motivated by the limitations of encoding raw images directly into a quantum circuit. To address this, we apply classical convolutional layers, which are an effective dimensionality reduction technique in reinforcement learning, as demonstrated in the seminal work of Mnih et al.~\cite{mnih_playing_2013}. This approach allows us to extract fewer but richer features from high-dimensional observations before encoding them in the PQC. As depicted in Figure~\ref{fig:hybrid-quantum-classical-model}, three convolutional layers together with a fully connected layer serve the purpose of dimensionality reduction and feature extraction. Together, these layers take the role of $L_{\boldsymbol{w}_{in}}$ in the notation established in Equation~\ref{eqn:hybrid-model}. The number of neurons at the output of the fully connected layer coincides with the number of encoding gates in the  PQC, as each neuron outputs a different latent feature. We refer to this layer as the \textit{pre-processing layer} and explain in more detail why we use only linear activations in this layer in Appendix~\ref{app:rescaling-latent-features}. The amount of features we can encode in the PQC is constrained by the limited number of qubits and encoding gates that we can handle computationally in our classical simulations. This in turn restricts the number of output neurons we can accommodate at the pre-processing layer. However, no such restriction exists for classical models conventionally used in RL, where the convolutional layers are typically followed directly by a wide non-linear processing layer. To allow for a fair assessment of the hybrid model's performance when compared to a classical counterpart, we introduce an artificial bottleneck in our classical reference models. We do so by adding an equivalent pre-processing layer with the same number of neurons as in the hybrid model. The exact specifications of the convolutional layers of both the hybrid and classical reference model as well as the architecture of the classical model are detailed in Appendix~\ref{app:classical-ref}.\\

The design of a PQC remains an area of active research, and a standard architecture has not yet been established. In this work, we choose to build on previously successful approaches for QRL in lower-dimensional observation spaces \cite{skolik_quantum_2022}: The architecture consists of multiple layers, each composed of one variational unitary (consisting of single-qubit $R_x$, $R_y$ and $R_z$ gates along with entangling $CZ$ gates) followed by one feature encoding unitary (a set of single-qubit $R_x$ rotations). This sequence of variational and encoding unitaries is repeated across all layers with a final variational unitary at the end of the circuit. The general structure of the PQC is shown in Figure~\ref{fig:pqc} and a more detailed description of the complete hybrid model is given in Appendix~\ref{app:detailed-architecture-hybrid-model}. The layerwise encoding of features leverages data re-uploading, which has been shown to greatly increase model expressivity \cite{perez-salinas_data_2020, schuld_effect_2021}. In our hybrid model, this technique takes on a unique form: instead of raw observations, the inputs to the PQC are trainable features produced by the preceding convolutional and fully connected layers. This adds to the model's flexibility and ability to learn complex functions. The output of the PQC are the expectation values of each qubit as obtained by local (i.e., qubit-wise) Pauli-$Z$ measurements. Unlike prior work by Skolik et al.~\cite{skolik_quantum_2022}, where each observable together with a single trainable parameter is assigned to a certain action in the environment, our model employs a more adaptive design. We use a fully connected linear post-processing layer, $L_{\boldsymbol{w}_{out}}$, to combine and rescale the PQC outputs, while also making the model compatible with action spaces larger than the number of available qubits.

\subsection{Quantum Q-learning}
\label{sec:quantum-q-learning}
Having established the architecture of our hybrid quantum-classical model, we now turn to its application in the RL framework, specifically within the Q-learning paradigm, which serves as the foundation for training and evaluating our model. RL is a field of ML where an agent interacts with an environment by taking actions to maximize cumulative rewards \cite{sutton_reinforcement_2018}. At a given time step $t$ the agent observes a state $s_t \in \mathcal{S}$ and selects an action $a_t \in \mathcal{A}$ based on a policy function $\pi(a|s)$ and receives feedback in the form of numerical rewards $r_t \in \mathcal{R}$, with the goal of learning an optimal policy $\pi^*$ through trial and error. The agent chooses actions so that to maximize the sum of discounted future rewards. In particular, it tries to maximize the expected discounted return $G_t := \sum_{k=t+1}^T \gamma^{k-t-1} r_k$, where the discount rate $0 < \gamma < 1$ determines the relative importance of short-term versus long-term rewards. The expected return when taking action $a$ while being in state $s$ and following the policy $\pi$ thereafter, is defined as the Q-function,

\begin{equation}
\label{eqn:q-value-function}
    Q_\pi(s, a) := \mathbb{E}_{\pi}[G_t|s_t=s, a_t=a].
\end{equation}

The optimal Q-function for a given state-value pair $(s, a)$ is defined as $Q^*(s, a) := \max_{\pi} Q_{\pi}(s, a)$. From the optimal Q-function $Q_*(s, a)$ an optimal policy $\pi_*$ can be derived: In state $s$, the agent should choose an action that maximizes the optimal Q-function, i.e

\begin{equation}
\label{eqn:optimal-policy}
    \pi^*(a|s)=\argmax\limits_a Q^*(s, a).
\end{equation}

This observation leads to the main objective of approximate Q-learning, where the goal is to learn an approximation $Q(s, a;\boldsymbol{\theta})$ of the optimal Q-function $Q^*(s, a)$ dependent on a set of parameters $\boldsymbol{\theta}$. A famous example is the deep Q-learning algorithm introduced in \cite{mnih_playing_2013}, where DNNs are used as Q-function approximators, labelled deep Q-networks (DQNs). Quantum Q-learning \cite{skolik_quantum_2022} is a variation of deep Q-learning, where the DQN is replaced by a PQC or a hybrid model as introduced in Section~\ref{sec:hybrid-model}. In this setting, the agent follows an $\varepsilon$-greedy policy, where $\varepsilon$ decays during training. At each time step $t$, the agent observes state $s_t$ and, with a probability $1-\varepsilon$, selects action $a_t$ corresponding to the highest estimated Q-value. Otherwise, it selects a random action. This behaviour balances exploration of the state space and exploitation of the current policy. Upon transition to state $s_{t+1}$ the agent obtains reward $r_t$ and the experience $e_t = (s_t, a_t,r_t, s_{t+1})$ is stored in a replay memory $D$. The replay memory $D$ is initially populated with $N$ experiences following a purely random policy, allowing the agent to sample a minibatch $B$ of transitions randomly at each time step, thereby breaking temporal correlations. For each sampled minibatch, the Q-network is trained by minimizing the following loss function,

\begin{equation}
\label{eqn:dqn-loss}
     \mathcal{L}(\boldsymbol{\theta}) = \Bigl(r_t + \gamma \max_{a'} \hat{Q}(s_{t+1}, a'; \boldsymbol{\theta}^-) - Q(s_t, a_t; \boldsymbol{\theta})\Bigl)^2, 
\end{equation}

where the term in parenthesis is the temporal difference (TD) error. The $\hat{Q}(s_{t+1}, a'; \boldsymbol{\theta}^-)$ indicates the use of a target network. It is a copy of the online network $Q(s_t, a_t; \boldsymbol{\theta})$, but its parameters $\boldsymbol{\theta}^-$ are updated less frequently. This technique avoids undesirable feedback loops and correlations between the target and the estimated Q-values, stabilizing the learning process. The loss function in Equation~\ref{eqn:dqn-loss} is minimized through gradient descent on the DQN weights, $\boldsymbol{\theta} \leftarrow \boldsymbol{\theta} - \alpha \nabla_{\boldsymbol{\theta}} \mathcal{L}(\boldsymbol{\theta})$, to refine the approximation of the Q-values. Every $C$ steps, the weights of the target network $\hat{Q}$ are updated to match the weights of the online network $Q$. Gradient computation for quantum circuits on quantum hardware is non-trivial due to quantum state collapse upon measurement; Appendix~\ref{sec:parameter-shift-rule} describes a method for obtaining gradients on real quantum computers. We refer to Appendix~\ref{app:training-procedure} for more details on the training procedure and give an overview of all hyperparameter settings investigated in Appendix~\ref{sec:hyperparam-settings}. Further, Appendix~\ref{app:atari-2600-environments} provides more information on the Atari environments and necessary pre-processing steps. Algorithm~\ref{alg:deep-q-learning} in Appendix~\ref{app:quantum-q-learning} shows a pseudo-code implementation of quantum Q-learning with experience replay and a target network. The complete implementation is available in this \href{https://github.com/Spiegeldondi/A-Hybrid-Quantum-Classical-Framework-for-Reinforcement-Learning-of-Atari-Games}{GitHub repository}.


\section{Model evaluation and analysis of design choices}
\label{sec:results}

In this section, our objectives are twofold: first, to demonstrate that our hybrid model can successfully learn in the Atari environments Pong and Breakout, and second, to analyze the impact of key design choices on its performance. We establish a baseline for both the quantum and classical models, showing that this baseline is sufficient to solve the Pong environment and achieves non-trivial performance in Breakout. Building on this, we systematically examine the effect of specific design choices — reward scaling and the latent feature space dimension — on the learning performance of both models.

\begin{figure}
 \def\figwidth{0.515\linewidth}
 \def\figheight{0.24\textheight} 

 \input{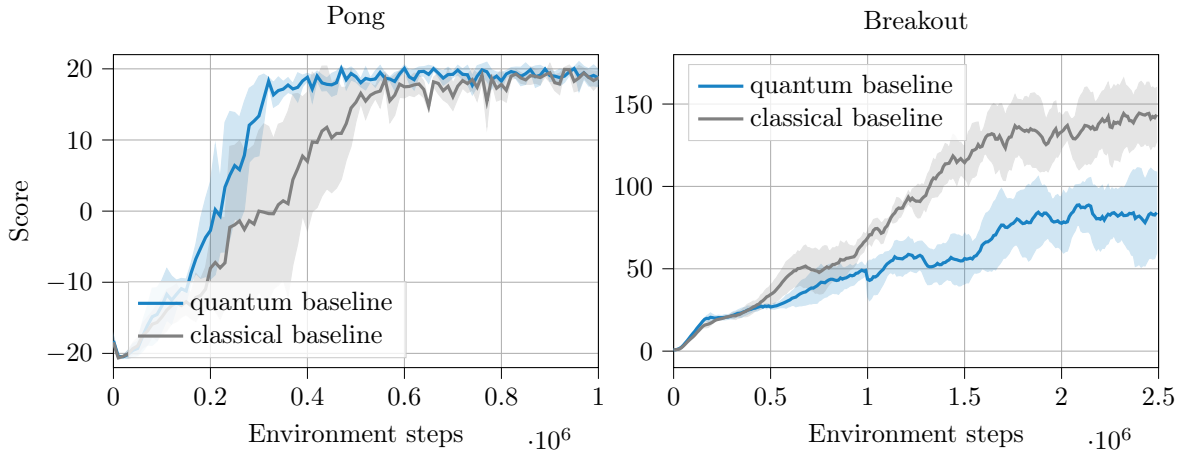}
 \vspace{-1em}
 \caption{\textbf{Rewards obtained during training for the hybrid and classical baseline models.} Shaded areas indicate the standard deviation of multiple runs. Left: The hybrid agent (blue) and the classical reference (grey) show differences in learning performance in Pong. In this environment, the hybrid agent appears to learn faster and more consistently across multiple runs. Right: Hybrid agent (blue) and classical reference (grey) in Breakout. Here, the hybrid model achieves a strong performance but shows a 41\% gap compared to the reference model.}
 \label{fig:scores-baseline}
\end{figure}

\subsection{Scores with baseline settings}
The scores in this setting (see 'q. baseline' and 'c. baseline' in Table~\ref{tab:parameter-settings} in Appendix~\ref{sec:hyperparam-settings}) serve as a baseline to which we compare the performance of agents with different design choices. Figure~\ref{fig:scores-baseline} illustrates the rewards obtained by the agent time-averaged over 10 episodes in the game of Pong (left) and 250 episodes in Breakout (right). As described in Section~\ref{sec:hyperparam-settings}, multiple runs were conducted for each environment using different random seeds. The solid lines indicate the average over these runs and the shaded region highlights the standard deviation. The results demonstrate that the hybrid model achieves a strong performance in both environments. Notably, in the game of Pong, the hybrid agent matches the classical reference in terms of final performance, achieving a mean reward of 20. Moreover, the hybrid model requires fewer environment steps - between 400,000 and 600,000 - to solve Pong, compared to around 800,000 steps needed by the classical agent. This result, however, does not indicate quantum advantage; it is likely influenced by statistical variability due to the limited number of runs. For the hybrid model, 4 out of 5 runs lead to successful learning behaviour, consistently achieving scores above 20. In contrast, only 4 out of 11 runs with the classical reference model showed successful learning. Notably, one of the successful classical runs exhibited a significant delay in reaching the optimal policy, which contributed to a lower average performance. The remaining runs were excluded from the analysis because they failed to learn at all, indicated by constant rewards of -21. In Breakout, the hybrid agent achieves a mean reward of approximately 84 after around 2 million episodes, which corresponds to a performance gap of about 41\% compared to the classical reference model. Despite this gap, the hybrid model demonstrates robust learning behaviour, closely following the trend of the classical agent. With additional design enhancements (see Figure~\ref{fig:less-bottleneck}), this gap can be significantly reduced to around 13\%. This demonstrates that the hybrid model is capable of achieving competitive performance and effectively solving complex reinforcement learning tasks, with clear potential for further optimization.

\begin{figure}
    \centering
    \includegraphics[width=0.85\textwidth]{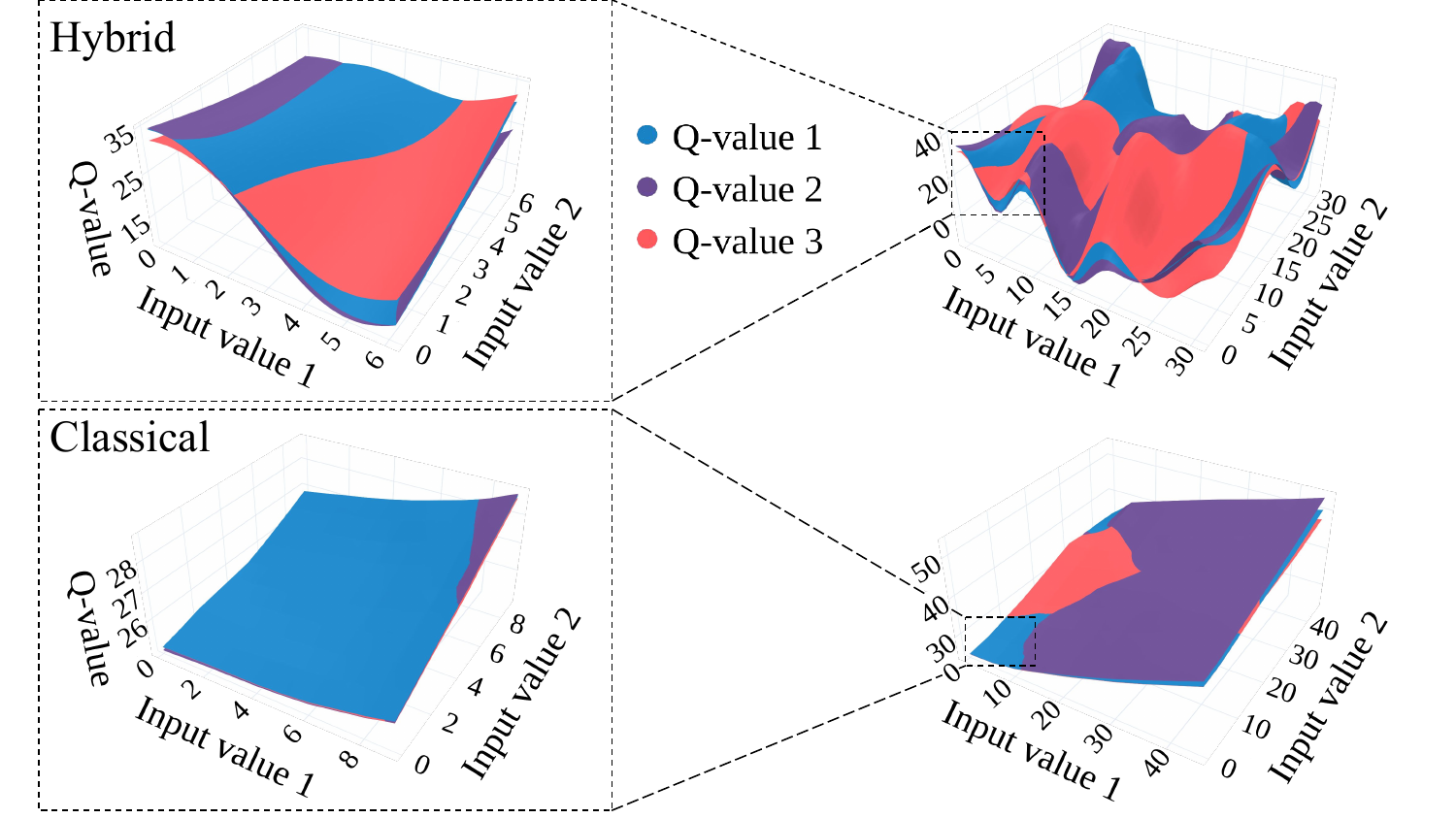}
    \caption{\textbf{Visualizing learned Q-values.} We plot the output of the hybrid and classical baseline models as a function of two randomly chosen inputs to the pre-processing layer, in place of the output generated by the convolutional layers in Figure~\ref{fig:hybrid-quantum-classical-model}. Left: Close-up of the Q-value surface, where the range of input values reflects a typical range as observed during an episode of Breakout. Right: The Q-value surface over an expanded region of the input space.}
    \label{fig:q-value-landscape}
\end{figure}

\subsection{Impact of reward scaling and post-processing layer learning rate}
\label{sec:results-reward-scaling}
Having established the potential for the hybrid model to learn effective policies in both testing environments, we now turn our attention to a more detailed analysis of how the Q-value estimates of the trained models compare. To do so, we generate \textit{Q-value surfaces} by plotting the three predicted Q-values as functions of two randomly selected inputs to the pre-processing layer, in place of the output of the convolutional layers (see Appendix~\ref{app:q-value-surface} for more details). Based on our observations, we investigate the impact of reward scaling and the learning rate of the post-processing layer on overall learning performance. Figure~\ref{fig:q-value-landscape} compares the Q-value surfaces of the hybrid and classical models. The left panels show surfaces for input values within a typical range observed during gameplay, while the right panels expand this range to better assess the surface shapes. The colours represent the Q-values corresponding to the three different actions.\\ 

A key observation is the difference in the shape of the Q-value surfaces. The hybrid model produces surfaces that are linear combinations of sinusoidal functions, reflecting the trigonometric nature of PQCs with rotational encoding, as analyzed in \cite{schuld_effect_2021}. In contrast, the classical model generates approximately piecewise linear surfaces. This difference has significant implications: We suspect that the oscillatory behaviour of the hybrid model's predictions increases the likelihood of assigning high Q-values to suboptimal actions. This can be understood by examining the number of crossings in the Q-value surfaces, which are more frequent in the hybrid model. Trigonometric functions, by nature, have a higher tendency for crossings due to their oscillatory behaviour, and small changes in their parameters can lead to numerous overlaps. This effect is likely even more pronounced in the higher-dimensional feature space produced by the convolutional layers. Our proposed solution to address this phenomenon is to upscale the rewards issued by the environment, thereby promoting greater separation between Q-values associated with optimal and suboptimal actions. This approach aims to provide more room for the oscillatory behaviour while reducing the chances of undesirable overlaps. While further investigation is required to fully understand this phenomenon, our results indicate that upscaling rewards benefits the hybrid model, leading to improved performance.\\

\begin{figure}
    \def\figwidth{0.515\linewidth}
    \def\figheight{0.24\textheight} 
    
    \input{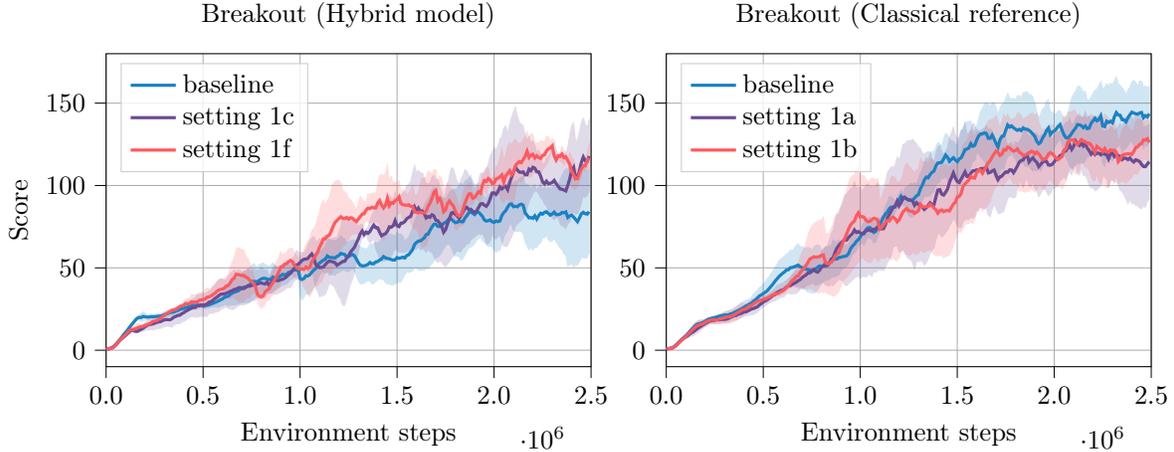}
    \vspace{-1em}
    \caption{\textbf{Analysis of the impact of reward rescaling in Breakout.} Left: The hybrid baseline model (blue) as well as a hybrid model trained with 10x scaled rewards and a final layer learning rate of 2.5e-2 (setting 1c) and a hybrid model trained with 100x reward scaling and a final layer learning rate of 2.5e-1 (setting 1f). Right: The classical reference model (blue) and a classic reference trained with 10x reward scaling and final layer learning rate of 2.5e-2 (setting 1a) and 100x reward scaling and final layer learning rate of 2.5e-1 (setting 1b). While the hybrid model benefits from the modifications, the performance of the classical model deteriorates.}
    \label{fig:reward-scaling}
\end{figure}

Closely related to the magnitude of the target Q-values is the learning rate of the post-processing layer following the PQC in the hybrid model, as it controls how quickly the model can adapt to certain target values. As discussed in Section~\ref{sec:hyperparam-settings} Table~\ref{tab:parameter-settings}, different combinations of reward scaling factors and learning rates of the pre-processing layer are tested. The best results are presented in Figure~\ref{fig:reward-scaling} (left) and compared to the hybrid baseline in blue. It is evident that the scaled rewards combined with a higher learning rate in the final layer of the hybrid model lead to significantly higher rewards (setting 1c and 1f). For both cases, the average reward clearly surpasses $100$ and gets close to a score of $120$. Results from literature \cite{skolik_quantum_2022} support the observation that higher learning rates in the post-processing part are beneficial. We find that the classical model does not benefit from reward scaling or higher learning rates, likely due to its inherently fewer Q-value crossings, which reduce the need for greater separation. Figure~\ref{fig:reward-scaling} (right) demonstrates that applying the settings beneficial for the hybrid model (settings 1a and 1b) to the classical model results in lower scores compared to the classical baseline.

\subsection{Influence of latent space dimension}
\begin{figure}
    \def\figwidth{0.515\linewidth}
    \def\figheight{0.24\textheight} 
    
    \input{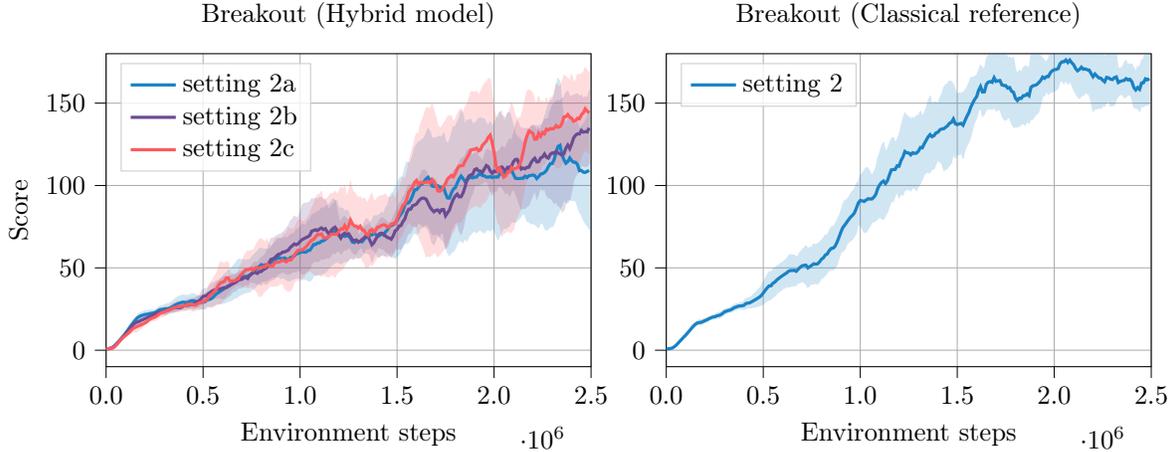}
    \vspace{-1em}
    \caption{\textbf{Analysis of the impact of latent space dimension and comparison of best-performing hybrid versus classical models in Breakout.} Left: Scores achieved by the hybrid model when the latent feature space dimension is increased by using more encoding gates in the PQC and expanding from 4 to 6 qubits in setting 2a. Settings 2b and 2c combine these relaxed constraints with reward scaling and higher learning rates in the post-processing. Right: The classical reference with increased latent space dimension also achieves noticeably higher scores when the constraint imposed by the bottleneck layer is loosened. Nonetheless, the classical-quantum performance gap is reduced to 13\% compared to the baseline results in Figure~\ref{fig:scores-baseline}.}
    \label{fig:less-bottleneck}
\end{figure}

A primary constraint on the performance of both the hybrid and classical model is the dimension of the latent space representation following the pre-processing layer. The number of latent features ultimately limits the information accessible by the PQC. In Figure~\ref{fig:less-bottleneck}, we demonstrate the impact of this latent space dimension on the hybrid model's performance and compare it to the classical reference. In configuration 2a, the PQC consists of 6 qubits (compared to 4 in the baseline model), each encoding 6 features per qubit (up from 4 in the baseline). This increases the latent space dimension from 16 to 36, more than doubling its size. Without additional modifications, this configuration achieves scores exceeding 100, a significant improvement over the baseline hybrid model. Building on the insights from Section~\ref{sec:results-reward-scaling}, we next examine configurations 2b and 2c, which combine the expanded latent space with reward scaling and increased learning rates in the post-processing layer. Specifically, configuration 2b incorporates a reward scaling factor of 10 and a learning rate of 2.5e-2, while configuration 2c uses a scaling factor of 100 and a learning rate of 2.5e-1. These adjustments lead to further improvements, with configuration 2c achieving a mean reward of nearly 150 within 2.5 million training steps, marking the highest performance achieved by the hybrid model under the given constraints. For comparison, the blue graph on the right side of Figure~\ref{fig:less-bottleneck} depicts the classical reference model with 36 features in the latent space. The improvement margin for this classical model over its baseline is similar to that of the hybrid model. It is important to note that excessively increasing the latent space dimension may negatively impact the hybrid model. Larger latent spaces require more qubits and deeper quantum circuits to encode all the features, increasing quantum circuit complexity. This added complexity can lead to trainability issues, such as the barren plateau phenomenon \cite{mcclean_barren_2018}, where gradients become exponentially small in the number of qubits. Exploring this regime numerically is computationally demanding and beyond the resources available for this project.


\section{Conclusion}
In this work, we have studied the performance of hybrid quantum-classical agents in RL environments characterized by a high-dimensional observation space. Specifically, we proposed a hybrid quantum-classical model based on PQCs for Q-function approximation and assessed its learning performance in the Atari game environments Pong and Breakout using a quantum approximate Q-learning framework. Our results demonstrate that hybrid models can learn effectively in these high-dimensional environments. This contrasts earlier work \cite{lockwood_playing_2021} that reported no learning capabilities in Atari environments due to the limited expressivity of the proposed hybrid model. Our research emphasizes the critical role of certain design choices, in particular the dimension of the latent feature space generated by the classical pre-processing layer, which ultimately controls the amount of information encoded in the PQC. Additionally, we identified how the magnitude of Q-values affects the hybrid model's learning performance when the learning rate of the classical post-processing layer is appropriately adjusted. Our findings suggest that with proper tuning, hybrid agents can closely match the performance of classical agents subjected to the same constraints in the latent space, highlighting the importance of fair benchmarking. The results presented contribute to our understanding of the interplay between quantum and classical components in hybrid models. We believe that this work marks a significant step towards the application of hybrid quantum-classical models in real-world RL scenarios. The performance of these models under the influence of noise, as present in today’s quantum hardware, poses an interesting open question. Future research could explore the robustness of the proposed architecture in noisy simulators and, ultimately, on real quantum devices. Another interesting research direction would be the exploration of different learning tasks where hybrid models could outperform fully classical models. Indeed, Atari games are known to be classically efficiently solvable, which leaves little room for a quantum advantage. One could instead explore environments where quantum enhancements are expected, as in quantum chemistry \cite{ostaszewski2021reinforcement} or combinatorial optimization \cite{patel2024reinforcement}. This research direction would require a deeper analysis of the interplay of the classical and quantum components of the hybrid model, as well as finding more task-specific PQC designs.

\section*{Acknowledgments}
The computational results presented have been achieved in part using the Vienna Scientific Cluster (VSC). SJ thanks the BMWK (EniQma) for their support.


\bibliographystyle{IEEEtran}
\bibliography{references}

\newpage


\appendix
\label{sec:appendix}


\section{Detailed architecture of the hybrid model}
\label{app:detailed-architecture-hybrid-model}

This section provides a detailed description of the hybrid quantum-classical model architecture complementing the overview in Section~\ref{sec:hybrid-model-architecture}. The hybrid model architecture comprises three main components (see also Figure \ref{fig:hybrid-quantum-classical-model} in the main text): first, classical convolutional layers together with a fully connected layer for dimensionality reduction and feature extraction; second, a PQC for processing the latent features; and third, a fully connected output layer for gathering and re-scaling expectation values.\\

The configuration of the convolutional layers in the hybrid model is identical to the classical reference model (see Appendix~\ref{app:classical-ref}). Following the convolutional layers is a fully connected pre-processing layer. This layer contains $n \times l$ neurons with linear activation, where $n$ is the number of qubits and $l$ is the number of PQC layers. This layer reduces the 3136 values produced by the convolutional layers to just 16 or 36 latent features, depending on the setting (see Table~\ref{tab:parameter-settings}). These features are directly encoded in the PQC.\\

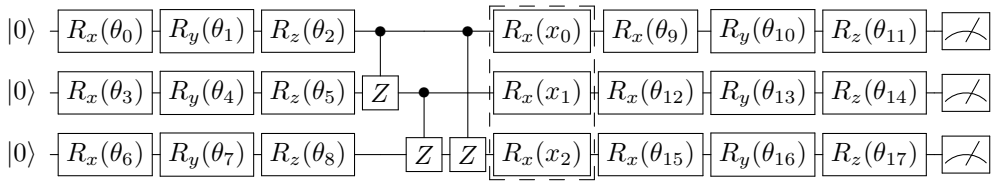
\begin{figure}[h]
\centering
\[
\Qcircuit @C=.3em @R=.7em {
\lstick{\ket{0}} & \gate{R_x(\theta_0)} & \gate{R_y(\theta_1)} & \gate{R_z(\theta_2)} & \ctrl{1} & \qw & \ctrl{2} & \gate{R_x(x_0)} & \gate{R_x(\theta_9)} & \gate{R_y(\theta_{10})} & \gate{R_z(\theta_{11})} & \qw & \meter\\
\lstick{\ket{0}} & \gate{R_x(\theta_3)} & \gate{R_y(\theta_4)} & \gate{R_z(\theta_5)} & \gate{Z} & \ctrl{1} & \qw & \gate{R_x(x_1)} & \gate{R_x(\theta_{12})} & \gate{R_y(\theta_{13})} & \gate{R_z(\theta_{14})} & \qw & \meter\\
\lstick{\ket{0}} & \gate{R_x(\theta_6)} & \gate{R_y(\theta_7)} & \gate{R_z(\theta_8)} & \qw & \gate{Z} & \gate{Z} & \gate{R_x(x_2)} & \gate{R_x(\theta_{15})} & \gate{R_y(\theta_{16})} & \gate{R_z(\theta_{17})} & \qw & \meter \gategroup{1}{8}{3}{8}{.3em}{--}\\
}
\]

\caption{\textbf{Quantum circuit diagram of the PQC with 3 qubits and a single layer.} A block of variational gates is followed by a set of entangling gates in a circular arrangement. The feature encoding block is highlighted by a dashed line. A second variational block comes after the encoding block. At the end of the circuit, local Pauli-$Z$ measurements are conducted.} 
\label{fig:pqc-structure} 
\end{figure}

The PQC at the core of the hybrid model follows a layer-wise structure inspired by prior work in quantum reinforcement learning \cite{skolik_quantum_2022}. Each layer comprises: 

\begin{enumerate}
    \item A variational block with an ${R}_x$-, ${R}_y$- and ${R}_z$-gate applied to each qubit in this order, each parameterized independently.
    \item An entanglement block with ${CZ}$-gates arranged in a circular topology, connecting qubits $1\rightarrow2$, $2\rightarrow3$, $3\rightarrow4$, ..., $n-1\rightarrow n$, $1\rightarrow n$ (${CZ}$-gates are symmetric).
    \item An encoding block where an ${R}_x$-gate is applied to each qubit, each encoding a distinct latent feature from the pre-processing layer.
\end{enumerate}

This PQC structure is repeated $l$ times, with a final variational block added at the end. The latent features are encoded layer-wise, where the first $n$ features are encoded in the first encoding block, the second $n$ features in the second encoding block and so forth. Consequently, the number of encoding gates in the circuit is $n \times l$. The number of variational parameters in the PQC is $(3 \times n) \times (l + 1)$. In contrast, the non-linear processing layer in the classical reference model contains $(n \times l \times 512) + 512$ trainable weights. In our baseline hybrid model (see setting "q. baseline" in Table \ref{tab:parameter-settings} in Appendix \ref{sec:hyperparam-settings}) with a 4-qubit PQC and 4 layers, the PQC has only 60 trainable parameters, compared to 8704 parameters in the corresponding non-linear layer of the classical reference model. Figure~\ref{fig:pqc-structure} depicts the PQC structure with $n=3$ and $l=1$. We initialize all variational parameters $\theta_i$ in the PQC according to the following distribution: $\theta^{init}_i \sim \mathcal{N}(0, \sigma^2)$ where $\sigma=0.01 \cdot \pi$.\\ 

The output of the PQC - expectation values from local Pauli-${Z}$ measurements - is passed to a classical post-processing layer. This layer consists of a fully connected linear layer where the number of neurons corresponds to the action space of the environment. This layer rescales the PQC's output to produce the final Q-values.\\


\section{Classical reference model} 
\label{app:classical-ref}
Here we introduce the classical reference model used as a fair baseline for evaluating the performance of the hybrid quantum-classical model. The architecture is based on the deep Q-network (DQN) from \cite{mnih_human-level_2015}, which we adopt without modification for the convolutional and fully connected layers. Specifically, the convolutional layers comprise three layers with the following parameters: The first layer consists of 32 filters of size $8 \times 8$ and a stride of 4, the second layer convolves 64 filters of size $4 \times 4$ and a stride of 2 and the third layer uses 64 filters of size $3 \times 3$ and a stride of 1. All convolutional layers use the non-linear \textit{rectifier} activation function (ReLU). The convolutional layers are followed by a fully connected layer with 512 neurons and ReLU activation, and a final output layer with linear activation.\\

\begin{figure}
    \centering
    \begin{subfigure}[b]{0.49\textwidth}
        \centering
        \includegraphics[width=1.0\textwidth]{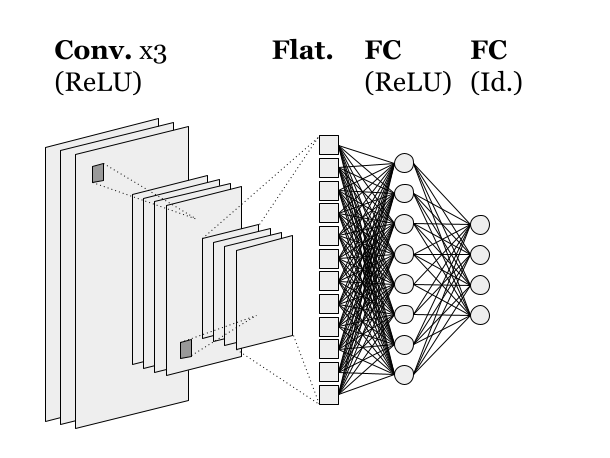}
        \label{fig:classic-mnih}
    \end{subfigure}
    \hfill
    \begin{subfigure}[b]{0.49\textwidth}
        \centering
        \includegraphics[width=1.0\textwidth]{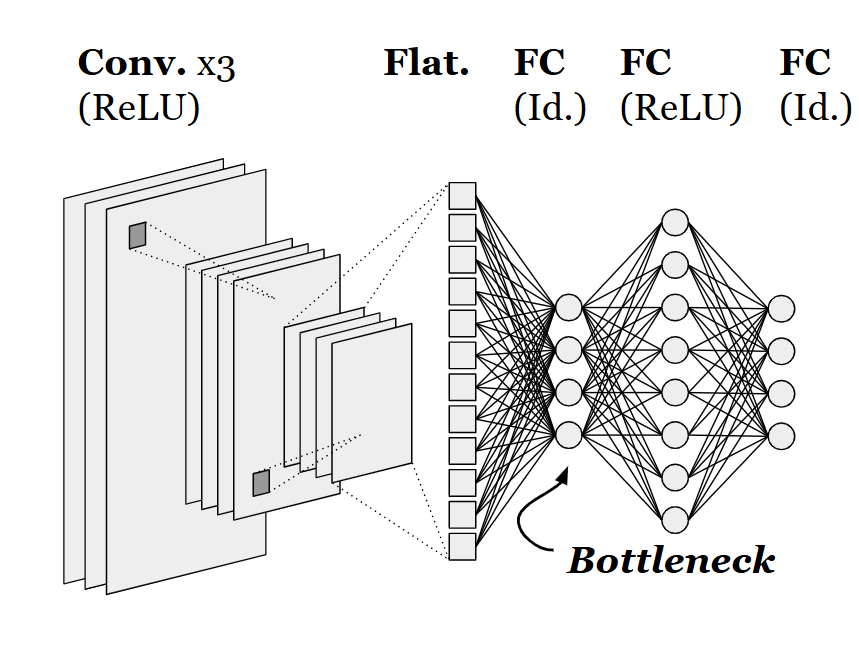}
        \label{fig:classic-ref}
    \end{subfigure}
    \caption{\textbf{Comparison between the original DQN architecture proposed in \cite{mnih_human-level_2015} (left) and our adapted version with a \textit{bottleneck} layer (right)}. The original model consists of three convolutional layers outputting 3136 features, a fully connected (FC) ReLU layer composed of 512 neurons, and a final output layer with linear activation. The adapted model incorporates a bottleneck layer (with 16 or 36 neurons) with linear activation after the convolutional layers, limiting the latent space dimension and mirroring the constraints of the hybrid quantum-classical model described.}
    \label{fig:classical-reference}
\end{figure}

In this work, this original architecture has been modified to allow a more direct comparison with the hybrid quantum-classical model. One of the key limitations of the hybrid model is its restricted latent space dimension, owing to the limited number of features that can be encoded in the PQC - as discussed in Section~\ref{sec:hybrid-model-architecture}. To account for this, a \textit{bottleneck layer} is introduced in the classical model between the convolutional layers and the fully connected ReLU layer. This bottleneck layer is implemented as a layer with a reduced number of neurons and linear activation, matching the dimensionality of the pre-processing layer in the hybrid model. The linear activation function is chosen to minimize the introduction of non-linearity, preserving the behaviour of the original architecture while introducing constraints in the latent space dimension. Figure~\ref{fig:classical-reference} compares the architecture proposed in \cite{mnih_human-level_2015} (left) to our adapted version with the bottleneck layer (right).\\

\begin{figure}
    \def\figwidth{0.515\linewidth}
    \def\figheight{0.24\textheight} 

     \input{graphics/figures/classical-mnih}

     \caption{\textbf{Classical model without bottleneck layer.} The left figure shows the rewards obtained in the game of Pong and the right figure shows the rewards obtained in Breakout. Without the bottleneck, the classical model converges faster and more consistently to the optimum in Pong and achieves significantly higher scores in Breakout.}
     \label{fig:classical-mnih}
 \end{figure}
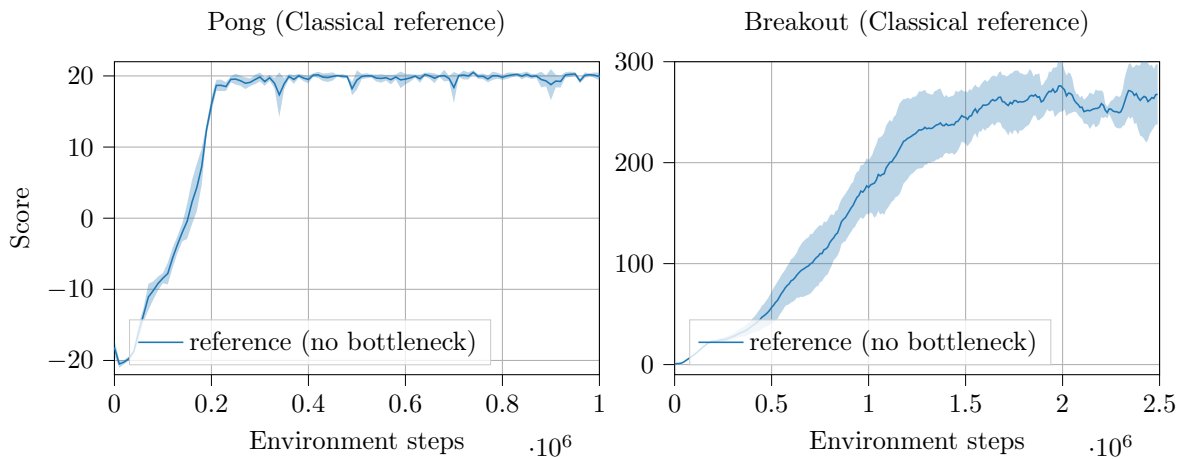

To assess the impact of the bottleneck layer, we compare the performance of the original model from \cite{mnih_human-level_2015} with our modified reference model. Figure~\ref{fig:classical-mnih} presents the rewards achieved by the classical model without the bottleneck constraint in both Pong and Breakout. The results demonstrate that the unconstrained model converges faster and more consistently in Pong and achieves significantly higher scores in Breakout. These findings highlight the performance limitations imposed by the bottleneck layer, which are also present in the hybrid model due to its architectural constraints.

\section{Atari 2600 environments}
\label{app:atari-2600-environments}
The Atari 2600 game environments of Pong and Breakout, accessible via OpenAI Gym \cite{brockman_openai_2016}, were selected as testbeds for evaluating the hybrid quantum-classical framework. These environments are well-established benchmarks in the field of reinforcement learning (RL) and have been extensively studied in classical settings. Both games meet the requirements for investigating the central research questions of this study, particularly due to their high-dimensional observation spaces represented by RGB images. Although Pong and Breakout share common elements in terms of their control structure, there are notable differences in their game dynamics. Breakout is generally considered a more challenging environment for RL agents to learn \cite{mnih_human-level_2015}, making it a suitable platform for testing advanced configurations of the hybrid model.\\

Pong simulates a two-player tennis-like game where the objective is to outscore a computer-controlled opponent. The player controls a paddle, which can be moved vertically to hit a ball back and forth, while the opponent aims to do the same. Points are awarded when the opponent either misses the ball or hits it out of bounds. The agent receives a reward of 1.0 for each point scored, with no penalties for losing a point. The game concludes when either player reaches a score of 21. The action space of Pong is limited to three relevant actions: "NOOP" (no operation), "RIGHT" (move up), and "LEFT" (move down), while actions such as "FIRE," "RIGHTFIRE," and "LEFTFIRE" are excluded since the ball is automatically launched at the beginning of each episode. This simplified action space allows for more focused training on the core mechanics of the game.\\

Breakout shares similar paddle-and-ball dynamics with Pong but introduces additional complexity by requiring the agent to destroy a wall of bricks. The player must control a paddle to bounce the ball upwards, with the goal of breaking bricks located at the top of the screen. Each brick destroyed yields a reward, with the value of the reward depending on the row of the brick: lower rows yield fewer points, while higher rows provide higher rewards. The maximum achievable score in Breakout is 432 points, reflecting the total points available across six rows of bricks. Similar to Pong, the action space in Breakout is constrained to "NOOP," "RIGHT," and "LEFT." An automatic firing mechanism is implemented to circumvent the need for the agent to learn to start the game manually. This reduces the complexity of the initial training phase by removing the "FIRE" action, which is only necessary to begin the game.

\subsection{Pre-processing of the raw environment state}
\begin{figure}
    \centering
    \includegraphics[width=0.75\textwidth]{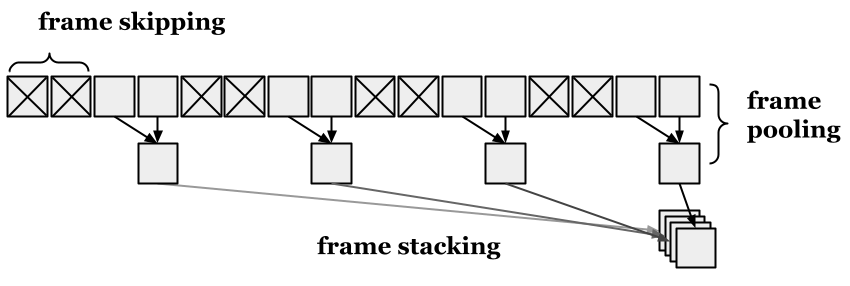}
    \caption{\textbf{The pre-processing pipeline illustrated on a sequence of game frames.} A rectangle represents a frame as produced by the game emulator. The frames with a cross are skipped and not shown to the agent. The last two frames of each sequence of four successive frames are combined using a frame pooling operation. Finally, the pooled frames are stacked to form a single new observation.}
    \label{fig:pre-processing-pipeline}
\end{figure}

\begin{figure}
    \centering
    \begin{subfigure}[b]{0.40\textwidth}
        \centering
        \includegraphics[width=0.6\textwidth]{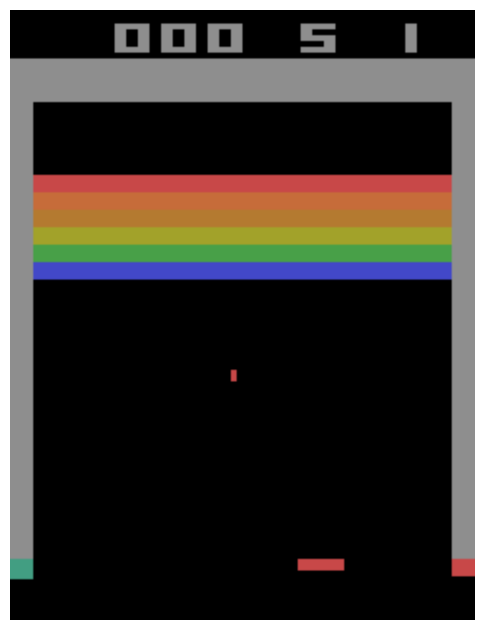}
        \label{fig:breakout_raw}
    \end{subfigure}
    \hfill
    \begin{subfigure}[b]{0.45\textwidth}
        \centering
        \includegraphics[width=0.6\textwidth]{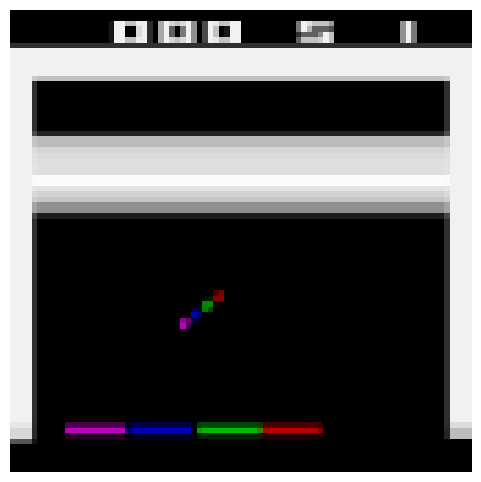}
        \label{fig:breakout_preprocessed}
    \end{subfigure}
    \caption{\textbf{Breakout observation space before (left) and after (right) pre-processing applied.} The four colours in the right image are for visualization purposes; each colour actually represents one of four distinct frames. The observation dimensions change from $210 \times 160 \times 3$ for the raw RGB image to $84 \times 84 \times 4$ for the cropped, grey-scaled and stacked observations.}
    \label{fig:atari_preprocessing}
\end{figure}

Before training the agent, several pre-processing steps are applied to the raw state representations of the game environments. These pre-processing steps, following the guidelines in \cite{machado_revisiting_2018}, are crucial for standardizing the evaluation of RL algorithms on Atari 2600 games and are independent of the pre-processing conducted by the hybrid model. The raw observation space for both environments consists of RGB images with a resolution of $210 \times 160 \times 3$, representing 210 pixels in height, 160 pixels in width, and three colour channels (red, green, and blue). The first step in the pre-processing pipeline involves converting the images to grayscale, reducing the number of colour channels from three to one. The images are then downsampled to a resolution of $84 \times 84$ pixels. To reduce the complexity of the RL problem, a technique known as \textit{frame skipping} is applied, where only every fourth frame of the game is used for decision-making. This means that the agent selects an action every four frames, with rewards accumulated over the skipped frames. Each of these decision points corresponds to a step in the environment. Additionally, a \textit{frame pooling} operation is applied to the last two frames of each four-frame sequence, combining them into a single frame. To introduce temporal information into the observation, four consecutive frames are stacked along the channel dimension, a process referred to as \textit{frame stacking}. Figure~\ref{fig:pre-processing-pipeline} illustrates this pre-processing pipeline and Figure~\ref{fig:atari_preprocessing} shows an instance of an environment observation in Breakout before and after applying the pre-processing. The final observation for the agent is thus represented as an $84 \times 84 \times 4$ tensor, where the four channels correspond to four consecutive frames rather than colour information. It is important to note that Atari 2600 environments are deterministic, meaning that each episode would evolve identically if restarted from the same seed. To mitigate the risk of the agent learning a fixed sequence of actions (open-loop control), stochasticity is introduced through $\epsilon$-greedy policies and random sampling from experience replay buffers. However, the strategy of applying random no-op actions at the start of the game, as suggested by \cite{mnih_human-level_2015}, is not used in this work, as it would lower the scores obtained by both hybrid and classical models, which is undesirable for the comparative analysis conducted in this study.

\section{Quantum Q-learning algorithm}
\label{app:quantum-q-learning}

In Section~\ref{sec:quantum-q-learning}, we introduced the quantum Q-learning algorithm based on the deep Q-learning algorithm. Algorithm~\ref{alg:deep-q-learning} presents a pseudo-code implementation following the notation established above.

\begin{algorithm}
    \SetKw{BreakFor}{break for}
    Initialize replay memory $D$ to capacity $N$
    
    Initialize action-value function $Q$ with random weights $\boldsymbol{\theta}$

    Initialize target action-value function $\hat{Q}$ with weights $\boldsymbol{\theta}^-=\boldsymbol{\theta}$
    
    \For{episode $=1$ \KwTo $M$}
    {
        Initialize sequence $s_1$
    
        \For{$t=1$ \KwTo $T$}
        {
            With probability $\varepsilon$ select random action $a_t$ otherwise select $a_t=\text{argmax}_a Q(s_t,a;\boldsymbol{\theta})$

            Execute action $a_t$ in emulator, observe reward $r_t$ and next state $s_{t+1}$

            Store transition $(s_t,a_t,r_t,s_{t+1})$ in $D$

            Sample random minibatch of size $|B|$ of transitions $(s_j,a_j,r_j,s_{j+1})$ from $D$ where $j$ is from the index set $B$

            \For{$j \in B$}
            {
                Set $y_j = \begin{cases}
                r_j, & \text{for terminal } s_{j+1} \\
                r_j + \gamma \max_{a'} \hat{Q}(s_{j+1},a';\boldsymbol{\theta}^-), & \text{for non-terminal } s_{j+1}
                \end{cases}$
            }
        
            Compute the loss for the minibatch: $\mathcal{L}(\boldsymbol{\theta}) = \frac{1}{|B|} \sum_{j \in B} (y_j - Q(s_j,a_j;\boldsymbol{\theta}))^2$

            Perform a gradient descent step: $\boldsymbol{\theta} \leftarrow \boldsymbol{\theta} - \alpha \nabla_{\boldsymbol{\theta}} \mathcal{L}(\boldsymbol{\theta})$
        
            Every $C$ steps, reset $\boldsymbol{\theta}^- = \boldsymbol{\theta}$
        }
    }
    \caption{Quantum Q-learning with experience replay and target network}
    \label{alg:deep-q-learning} 
\end{algorithm}


\section{Training procedure and evaluation metrics}
\label{app:training-procedure}
We evaluate the hybrid quantum-classical model within a reinforcement learning framework using quantum Q-learning with a target network and experience replay, following the approach of Mnih et al.~\cite{mnih_human-level_2015}. Training starts with a random policy for 20,000 environment steps to populate the replay buffer. After accumulating 100,000 $(s_t, a_t, r_t, s_{t+1})$ transitions, the replay buffer operates on a first-in, first-out (FIFO) basis with uniform sampling. After this
initial warm-up phase, the predictions of the hybrid agent are used to derive a policy. The online model is trained every 4 steps, and the target network is updated every 8,000 steps. In the baseline setting (see 'q. baseline' in Table~\ref{tab:parameter-settings}), we minimize the loss function (Equation~\ref{eqn:dqn-loss}) using the Adam optimizer with a learning rate of $2.5 \times 10^{-4}$ for both classical and quantum components. An $\epsilon$-greedy policy is employed with $\epsilon$ linearly decaying from 1 to 0.01 over 250,000 steps. The discount factor $\gamma$ is set to 0.99. Training spans 1 million steps in Pong and 2.5 million steps in Breakout. The main evaluation metric employed for assessing the agents in this study is the total undiscounted reward obtained in each episode. These rewards exhibit significant variance across different game episodes, due to the $\epsilon$-greedy policy and random sampling from the replay memory. For evaluation purposes, the rewards are temporally averaged, where we take 10 episodes in Pong and 250 episodes in Breakout. Following the recommendations in \cite{machado_revisiting_2018}, no evaluation runs where the agent fully exploits its learned policy (without any parameter updates) are conducted.


\section{Numerical experiments and hyperparameter settings} 
\label{sec:hyperparam-settings}
This section provides a brief overview of the hyperparameters we investigate and the software frameworks used.
Table~\ref{tab:parameter-settings} outlines the hyperparameter values tested and indicates the number of runs conducted for each setting to ensure statistically reliable results and mitigate the effects of randomness. The learning rate of the post-processing layer is identified as critical for Q-learning, referring to results presented in \cite{skolik_quantum_2022}, with settings 1a to 1f in Table~\ref{tab:parameter-settings} exploring various learning rates. Alongside adjusting the learning rate of the post-processing layer, we upscale rewards to mitigate the impact of the hybrid model's oscillatory Q-value behaviour and reduce undesirable overlaps (see Section~\ref{sec:results-reward-scaling}). Additionally, as the number of latent features, determined by the number of neurons in the classical pre-processing layer, ultimately limits the information accessible to the PQC, we compare the baseline to settings with higher latent space dimensions (2a to 2c). The values of all hyperparameters are tested through an informal search conducted in Breakout. We refrain from a systematic grid search due to significant computational requirements. The settings in which the hybrid model performed best were also tested with the classical model for reference. All numerical experiments are conducted using the TensorFlow Quantum \cite{broughton_tensorflow_2021} module for noise-free quantum circuit simulations and TensorFlow Agents \cite{TFAgents} for RL-related algorithms. The game environments, Pong and Breakout, are provided by OpenAI Gym \cite{brockman_openai_2016}. 

\begin{table}[ht]
\centering
\begin{tabular}{ | c || c | c | c | c | }
    \hline
    \multicolumn{5}{|c|}{Hyperparameter settings} \\
    \hline
    \textbf{Setting} & \textbf{Learning rate} & \textbf{Reward scaling} & \textbf{Latent features} & \textbf{Number of runs}\\
    \hline
    q. baseline & 2.5e-4 & - & 16 & 5 (4 in Pong)\\
    \hdashline
    quantum 1a & 2.5e-3 & - & 16 & 3\\
    quantum 1b & 2.5e-3 & 10x & 16 & 3\\
    quantum 1c & 2.5e-2 & 10x & 16 & 3\\
    quantum 1d & 2.5e-2 & 100x & 16 & 3\\
    quantum 1e & 2.5e-1 & 10x & 16 & 3\\
    quantum 1f & 2.5e-1 & 100x & 16 & 3\\
    \hdashline
    quantum 2a & 2.5e-4 & - & 36 & 5\\
    quantum 2b & 2.5e-2 & 10x & 36 & 5\\
    quantum 2c & 2.5e-1 & 100x & 36 & 5\\
    \hline
    c. baseline & 2.5e-4 & - & 16 & 5 (4 in Pong)\\
    \hdashline
    classical 1a & 2.5e-2 & 10x & 16 & 5\\
    classical 1b & 2.5e-1 & 100x & 16 & 5\\
    \hdashline
    classical 2 & 2.5e-4 & - & 36 & 5\\
    \hline
\end{tabular}
\caption{\textbf{A summary of the numerical experiment settings.} The first column provides labels to identify each hyperparameter setting. Here, 'quantum' or 'q.' refers to the hybrid model and 'classical' or 'c.' to the classical reference. 'Learning rate' is the learning rate of the classical post-processing layer. 'Reward scaling' indicates by which factor rewards are up-scaled and 'Latent features' resembles the latent space dimension in terms of the number of features. For the classical reference, only those 'Reward scaling' settings are chosen, that performed best for the hybrid model.}
\label{tab:parameter-settings}
\end{table}


\begin{figure}
    \def\figwidth{0.515\linewidth}
    \def\figheight{0.24\textheight} 
    
    \input{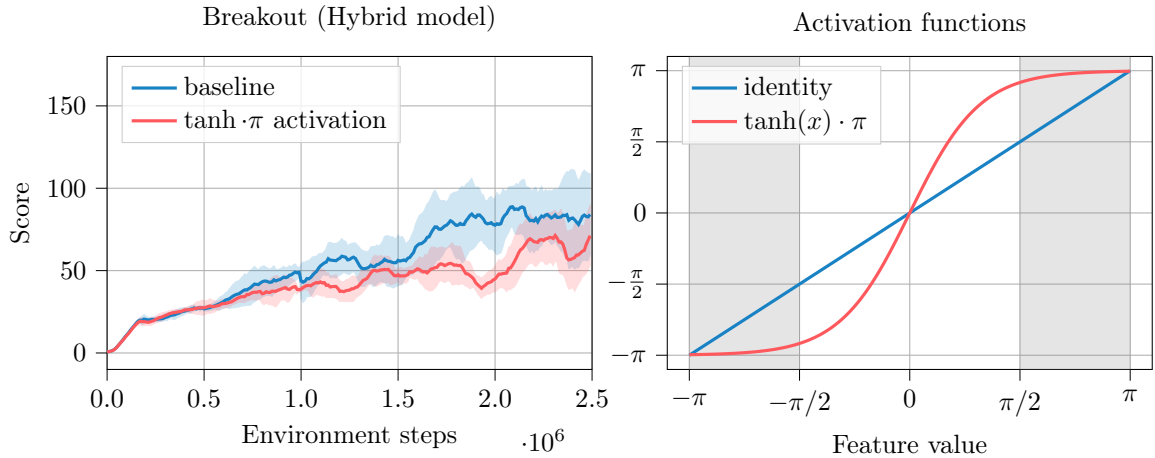}
    
\caption{\textbf{Performance of the hybrid model depending on the pre-processing layer activation function.} Left: Scores achieved by the baseline hybrid model (baseline) and a hybrid model with $\tanh\cdot\pi$ activation function in the pre-processing layer ($\tanh\cdot\pi$) in Breakout. The $\tanh\cdot\pi$ activation in the pre-processing layer leads to worse performance. Right: Comparison of the linear (identity) function (blue) and a $tanh$ function scaled by a factor of $\pi$ (red). Applying $\tanh\cdot\pi$ to feature values in the shaded area $\pm(0.5\pi, \pi)$, maps them to a very narrow region in the image of $\tanh\cdot\pi$, possibly resulting in information loss.}
    \label{fig:scores-tanh}
\end{figure}

\begin{figure}
    \def\figwidth{0.515\linewidth}
    \def\figheight{0.24\textheight} 
    
    \input{graphics/figures/feature-distributions}
    
    \caption{\textbf{Distributions of individual latent feature values during training.} Left: The feature values of the hybrid model remain within the range of $\pm\pi$. Right: The feature values in the classical reference model exceed this range, indicating that the classical pre-processing layers in the hybrid model adapt to the rotational encoding constraints of the PQC.}
    \label{fig:feature-distributions}
\end{figure}
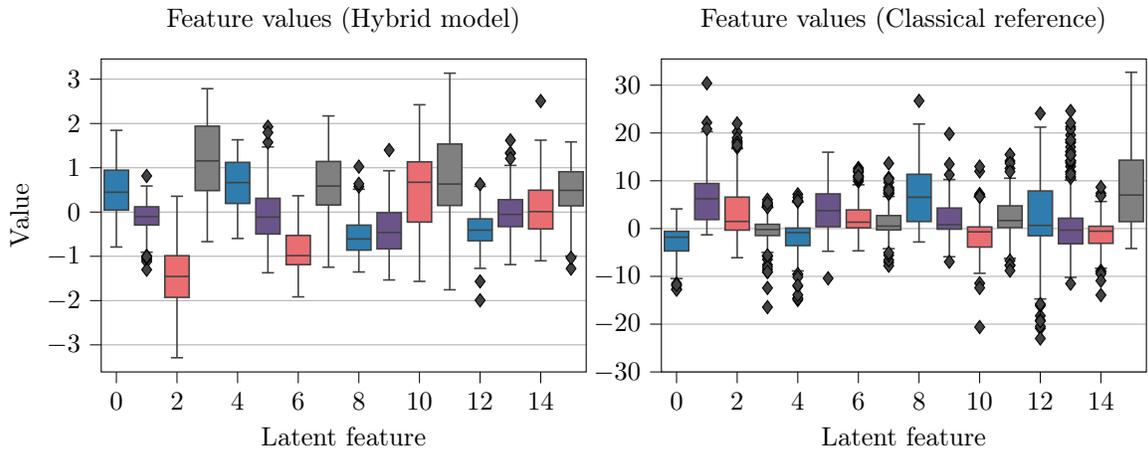

\section{Effect of rescaling latent features}
\label{app:rescaling-latent-features}
A key design choice in the hybrid model architecture is the use of a linear activation function in the pre-processing layer preceding the PQC. This decision aims to preserve the original latent features and avoid introducing any transformations that might overshadow the learning process within the quantum circuit. However, this approach does not impose any constraint on the magnitude of the latent features produced by the pre-processing layer. Since the PQC encodes features as rotation angles, it is essential to ensure that the values of the latent features do not exceed a range of $2\pi$, as otherwise two vastly different features might be mapped to the same value due to the periodicity of the encoding scheme. One common strategy to enforce this constraint is to apply a scaling activation function, such as $\tanh\cdot\pi$, which restricts the output of the pre-processing layer to values within the desired range. Figure~\ref{fig:scores-tanh} (left) compares the performance of the baseline hybrid model introduced in Section~\ref{sec:hybrid-model-architecture} with the linear activation function to a hybrid model employing the $\tanh\cdot\pi$ activation function. The results demonstrate that the baseline model achieves better performance, suggesting that rescaling latent features is not necessary. A possible explanation for the poor performance of the scaling activation function lies in the distortion introduced by the $\tanh\cdot\pi$ function. As shown in Figure~\ref{fig:scores-tanh} (right), the feature values near $\pm\pi$ are mapped to a narrow range, potentially diminishing the influence of large latent features (in absolute terms) and thus reducing the amount of meaningful information encoded in the PQC.  Further analysis of the latent feature distributions, as shown in Figure~\ref{fig:feature-distributions} (left), suggests that a linear activation function in the pre-processing layer is sufficient for maintaining the necessary range of values. The distribution of feature values, sampled every 10,000 environment steps during training, indicates that the majority of the values remain well within the range of $\pm 2\pi$, with most falling within $\pm\pi$. This observation implies that the classical layers preceding the PQC, particularly the pre-processing layer, naturally learn to regulate the magnitudes of latent features during training, without requiring an explicit scaling activation function. In comparison, Figure~\ref{fig:feature-distributions} (right) shows the distribution of feature values in the classical reference model, which does not have the same rotational encoding constraints. As expected, the magnitudes of the latent features in the classical model exceed the range observed in the hybrid model. This further supports the hypothesis that the classical pre-processing components of the hybrid model adapt to respect the constraints imposed by the quantum encoding, without the need for additional feature rescaling.


\begin{figure}
    \centering
    \includegraphics[width=0.77\textwidth]{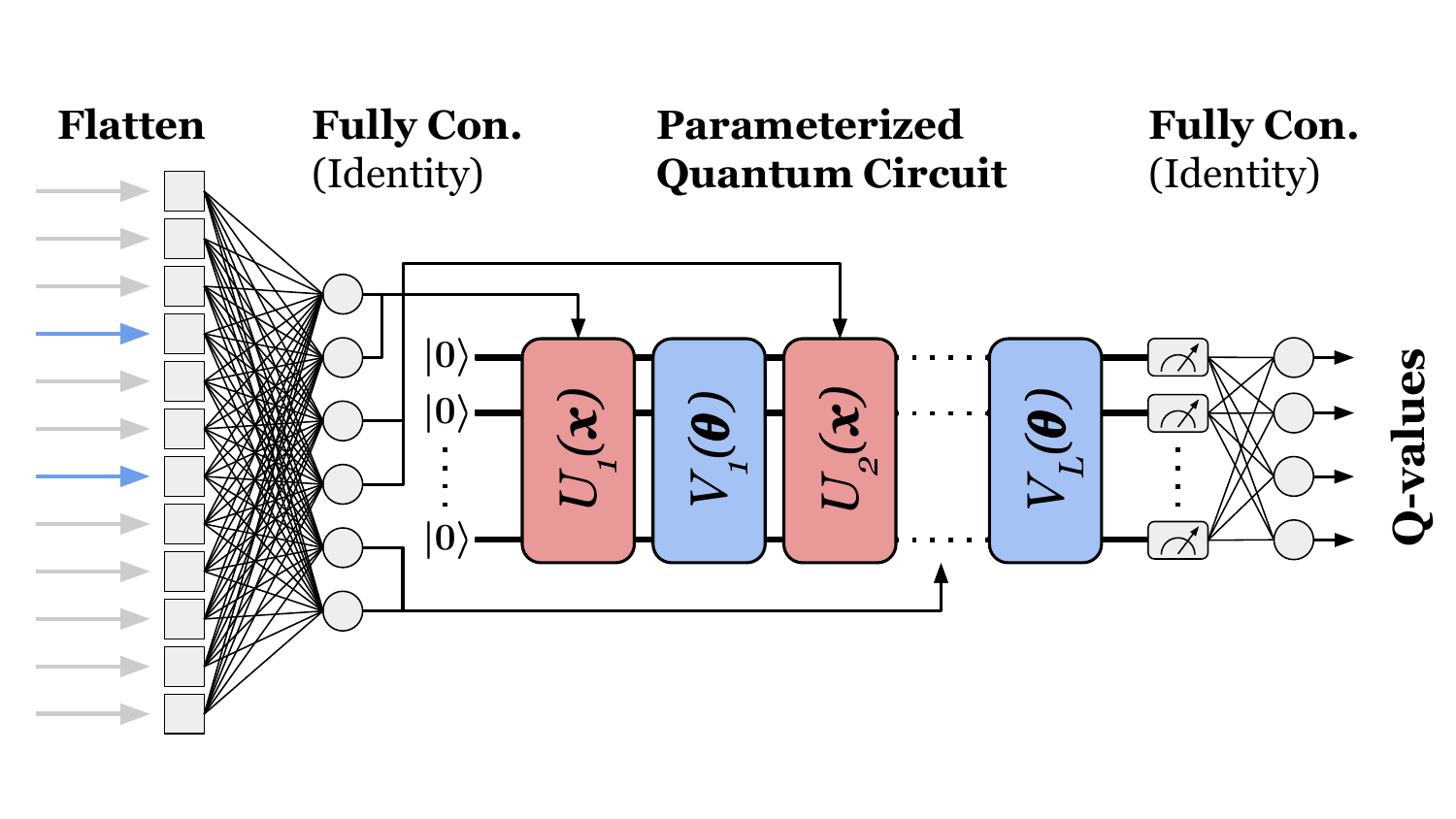}
    \vspace{-1.5em}
    \caption{\textbf{A truncated hybrid model without convolutional layers}. This model is used to plot predicted Q-values as a function of two randomly chosen inputs to the pre-processing layer, in place of the output of the convolutional layers. All except for two inputs (blue arrows) to the truncated model are kept constant. The Q-value predictions are plotted as a function of these two inputs resulting in a Q-value surface plot.}
    \label{fig:q-value-landscape-model}
\end{figure}

\section{Plotting Q-value surfaces}
\label{app:q-value-surface}
Here we explain how the Q-value surface plots presented in Section~\ref{sec:results-reward-scaling} are generated for both the hybrid and classical model. We analyze the predicted Q-values as a function of inputs to the pre-processing layer. The output vectors of the convolutional layers (after flattening) are logged for each step of a single episode of Breakout, resulting in a collection of 3136-dimensional vectors. From these, we compute an average output vector, providing a representative baseline for the convolutional network's outputs during the episode.\\

Next, a copy of the model is created, excluding the convolutional layers. Since the full 3136-dimensional input space is too large to visualize directly, we select two features (input nodes of the truncated model) at random to reduce the dimensionality to a more interpretable two-dimensional space (see Figure~\ref{fig:q-value-landscape-model}). This allows us to visualize the predicted Q-values as surface plots, providing an intuitive means to analyze how variations in selected features affect the model's output.\\

For the two selected features, multiple values are fed into the pre-processing layer, varying over a range observed during the episode. All other input features are fixed to their respective average values computed earlier. This approach generates three Q-value surfaces (corresponding to the three actions in Breakout) as functions of the two selected features.


\section{Estimating gradients of quantum computations}
\label{sec:parameter-shift-rule}

As discussed in Section~\ref{sec:quantum-q-learning}, training quantum machine learning (QML) models involves minimizing a cost or loss function $\mathcal{L}(\boldsymbol{\theta})$, which measures the discrepancy between computed expectation values $f_{\boldsymbol{\theta}}(\boldsymbol{x})$ and target values. Like in classical machine learning, this optimization task can be tackled with gradient-based methods like gradient descent and its variations in a hybrid quantum-classical manner. Once the gradient of the cost function with respect to the variational gate parameters $\nabla_{\boldsymbol{\theta}} \mathcal{L}(\boldsymbol{\theta})$ is known, the classical computer can update the parameters of the PQC in the direction of smaller cost by performing a gradient descent step. In classical machine learning, gradients are obtained via \textit{backpropagation}, a form of automatic differentiation (autodiff), which uses intermediate values to efficiently calculate the gradient of a function with a runtime similar to the execution of the function itself. This dependency on intermediate information makes it difficult to apply backpropagation, or autodiff in general, to quantum computations, as the collapse of the quantum state upon measurement prevents the straight-forward reuse of information \cite{mari_estimating_2021, abbas_quantum_2023}. This fundamental difference requires alternative approaches for gradient computation in quantum machine learning, such as the \textit{parameter-shift rule} \cite{mitarai_quantum_2018, schuld_evaluating_2019, mari_estimating_2021}.\\ 

The parameter-shift rule exploits the trigonometric nature of PQCs as functions of their parameters. As a function of a single variational parameter $\theta_i$ with all other parameters $\theta_j \in \boldsymbol{\theta}, j \neq i$, and the inputs $\boldsymbol{x}$ fixed, the quantum model as defined in Equation~\ref{eqn:determ-qml-model} is of the form

\begin{equation}
\label{eqn:pqc-single-parameter}
   f_{\theta_i} = a\cdot\sin(\theta_i + b) + c,
\end{equation}

where the constants $a$, $b$ and $c$ are determined by the fixed part of the circuit. Functions of this type adhere to the parameter-shift rule, which allows the exact computation of the derivative with respect to the parameter $\theta_i$ using two evaluations of the function with a shift in the parameter. Specifically, for a given parameter $\theta_i$ in a quantum circuit, the gradient of the expectation value $f_{\theta_i} := \braket{\mathcal{M}}_{\boldsymbol{\theta_i}}$ can be evaluated as:

\begin{equation}
\label{eqn:parameter-shift-rule}
\frac{\partial f_{\theta_i}} {\partial \theta_i} = \frac{f_{\theta_i + \pi/2} - f_{\theta_i - \pi/2}}{2} = a\cdot\cos(\theta_i + b).
\end{equation}

This applies to an expectation value as in Equation~\ref{eqn:determ-qml-model} - the deterministic quantum machine learning model $f_{\boldsymbol{\theta}}$. It is important to note that Equation~\ref{eqn:parameter-shift-rule} is not a finite difference approximation but gives the exact value of the gradient - if the expectation values $\braket{\mathcal{M}}_{\boldsymbol{\theta}}$ were known exactly. In that sense, an estimation of the exact gradient can be computed via Equation~\ref{eqn:parameter-shift-rule} using 2$s$ circuit evaluations per parameter $\theta_i$, where $s$ is the number of shots to estimate an expectation value. When \emph{simulating} quantum computations on classical computers, the situation is however different: The gradients of simulated quantum computations can be evaluated without the need for the parameter-shift rule. Since the quantum state of the system is completely known during all times of the computation, classical automatic differentiation can be applied directly.\\

\end{document}

%% file: graphics/figures/classical-mnih.tex
\begin{tikzpicture}

\definecolor{darkgray176}{RGB}{176,176,176}
\definecolor{lightgray204}{RGB}{204,204,204}
\definecolor{steelblue31119180}{RGB}{31,119,180}

\begin{groupplot}[group style={group size=2 by 1}]
\nextgroupplot[
height=\figheight,
legend cell align={left},
legend style={
  fill opacity=0.8,
  draw opacity=1,
  text opacity=1,
  at={(0.03,0.03)},
  anchor=south west,
  draw=lightgray204
},
tick align=outside,
tick pos=left,
title={Pong (Classical reference)},
width=\figwidth,
x grid style={darkgray176},
xlabel={Environment steps},
xmajorgrids,
xmin=0, xmax=1000000,
xtick style={color=black},
y grid style={darkgray176},
ylabel={Score},
ymajorgrids,
ymin=-22, ymax=22,
ytick style={color=black}
]
\path [fill=steelblue31119180, fill opacity=0.3]
(axis cs:0,-17.0642374527743)
--(axis cs:0,-19.3107625472257)
--(axis cs:10000,-21.0338539126016)
--(axis cs:20000,-20.5111380785565)
--(axis cs:30000,-20.0281088913246)
--(axis cs:40000,-19.0811951712194)
--(axis cs:50000,-17.0598386976874)
--(axis cs:60000,-14.5246709956611)
--(axis cs:70000,-12.9139874931603)
--(axis cs:80000,-11.4718712349343)
--(axis cs:90000,-10.1557522622967)
--(axis cs:100000,-9.15429760729074)
--(axis cs:110000,-9.30695136998535)
--(axis cs:120000,-6.98810827610626)
--(axis cs:130000,-4.90208152801713)
--(axis cs:140000,-3.25635670287395)
--(axis cs:150000,-2.89722066167242)
--(axis cs:160000,-0.834159854365806)
--(axis cs:170000,1.00448060833598)
--(axis cs:180000,4.69608144217557)
--(axis cs:190000,11.7)
--(axis cs:200000,14.8905189949791)
--(axis cs:210000,17.8959781774559)
--(axis cs:220000,17.8707146028927)
--(axis cs:230000,17.8876275643042)
--(axis cs:240000,19.0927095883552)
--(axis cs:250000,18.780259784083)
--(axis cs:260000,18.6060598077726)
--(axis cs:270000,18.1492427717543)
--(axis cs:280000,18.5614835192866)
--(axis cs:290000,18.8404547020864)
--(axis cs:300000,19.2155711229775)
--(axis cs:310000,18.5958477013203)
--(axis cs:320000,19.3346688068541)
--(axis cs:330000,18.2422824555124)
--(axis cs:340000,14.196034242418)
--(axis cs:350000,18.1384226894136)
--(axis cs:360000,19.5758939989058)
--(axis cs:370000,18.8876275643042)
--(axis cs:380000,19.7388619214435)
--(axis cs:390000,19.2584119566836)
--(axis cs:400000,18.9661460873984)
--(axis cs:410000,19.8762531407233)
--(axis cs:420000,19.7112517806304)
--(axis cs:430000,19.2847454303756)
--(axis cs:440000,19.1570396452846)
--(axis cs:450000,19.3402992218237)
--(axis cs:460000,19.938196601125)
--(axis cs:470000,19.6298437881284)
--(axis cs:480000,19.7129171306613)
--(axis cs:490000,17.2832595899081)
--(axis cs:500000,18.058663657806)
--(axis cs:510000,19.5550337853263)
--(axis cs:520000,19.6618540238416)
--(axis cs:530000,19.7654792120088)
--(axis cs:540000,19.2152320142584)
--(axis cs:550000,19.0479202710604)
--(axis cs:560000,19.3468871125851)
--(axis cs:570000,18.806512902144)
--(axis cs:580000,19.5)
--(axis cs:590000,18.2827382470071)
--(axis cs:600000,18.8066965626341)
--(axis cs:610000,19.1893928678593)
--(axis cs:620000,19.7584936490539)
--(axis cs:630000,19.0212081548605)
--(axis cs:640000,19.9550510257217)
--(axis cs:650000,19.5869646133016)
--(axis cs:660000,18.9880684167983)
--(axis cs:670000,19.8450961894323)
--(axis cs:680000,19.2329932680816)
--(axis cs:690000,19.3)
--(axis cs:700000,16.193973865356)
--(axis cs:710000,19.8438447187191)
--(axis cs:720000,19.4990007987226)
--(axis cs:730000,19.5050252531694)
--(axis cs:740000,20.2084524052577)
--(axis cs:750000,19.5905227959935)
--(axis cs:760000,19.5468871125851)
--(axis cs:770000,19.0934659928935)
--(axis cs:780000,19.4489140689099)
--(axis cs:790000,19.6044349039685)
--(axis cs:800000,19.4464466094067)
--(axis cs:810000,19.5756949811097)
--(axis cs:820000,19.8298437881284)
--(axis cs:830000,20.0464642892864)
--(axis cs:840000,19.5201648650335)
--(axis cs:850000,19.835288568297)
--(axis cs:860000,19.5688048287806)
--(axis cs:870000,19.7)
--(axis cs:880000,18.3932028189411)
--(axis cs:890000,18.5221825406948)
--(axis cs:900000,16.593973865356)
--(axis cs:910000,18.3879402500395)
--(axis cs:920000,18.3783152889062)
--(axis cs:930000,19.7659427126066)
--(axis cs:940000,19.848337020667)
--(axis cs:950000,20.0069048676309)
--(axis cs:960000,18.9608835008437)
--(axis cs:970000,19.7062996062994)
--(axis cs:980000,19.9829713563033)
--(axis cs:990000,19.7062996062994)
--(axis cs:1000000,19.3847454303756)
--(axis cs:1010000,19.9137525100503)
--(axis cs:1020000,20.1450490243204)
--(axis cs:1030000,19.5699264745632)
--(axis cs:1040000,18.6550167787124)
--(axis cs:1050000,18.5800943452208)
--(axis cs:1060000,19.9450490243204)
--(axis cs:1070000,17.9440643959029)
--(axis cs:1080000,19.0602632099846)
--(axis cs:1090000,19.9)
--(axis cs:1090000,20.6)
--(axis cs:1090000,20.6)
--(axis cs:1080000,20.3897367900154)
--(axis cs:1070000,20.0559356040971)
--(axis cs:1060000,20.4549509756796)
--(axis cs:1050000,20.1699056547792)
--(axis cs:1040000,20.1449832212876)
--(axis cs:1030000,20.9300735254368)
--(axis cs:1020000,20.6549509756796)
--(axis cs:1010000,20.5362474899497)
--(axis cs:1000000,20.4652545696244)
--(axis cs:990000,20.4937003937006)
--(axis cs:980000,20.3670286436967)
--(axis cs:970000,20.4937003937006)
--(axis cs:960000,19.6391164991563)
--(axis cs:950000,20.5430951323691)
--(axis cs:940000,20.601662979333)
--(axis cs:930000,20.5340572873934)
--(axis cs:920000,20.0716847110938)
--(axis cs:910000,20.1620597499605)
--(axis cs:900000,20.956026134644)
--(axis cs:890000,20.0778174593052)
--(axis cs:880000,20.6067971810589)
--(axis cs:870000,20.3)
--(axis cs:860000,20.2811951712194)
--(axis cs:850000,20.614711431703)
--(axis cs:840000,20.4298351349665)
--(axis cs:830000,20.4035357107136)
--(axis cs:820000,20.4701562118716)
--(axis cs:810000,20.4743050188903)
--(axis cs:800000,20.1535533905933)
--(axis cs:790000,20.4455650960315)
--(axis cs:780000,20.6010859310901)
--(axis cs:770000,20.0565340071065)
--(axis cs:760000,20.3531128874149)
--(axis cs:750000,20.2594772040065)
--(axis cs:740000,20.7915475947423)
--(axis cs:730000,20.4949747468306)
--(axis cs:720000,20.7509992012774)
--(axis cs:710000,20.2561552812809)
--(axis cs:700000,20.556026134644)
--(axis cs:690000,20)
--(axis cs:680000,20.8670067319184)
--(axis cs:670000,20.1049038105677)
--(axis cs:660000,20.3619315832017)
--(axis cs:650000,20.4630353866984)
--(axis cs:640000,20.4449489742783)
--(axis cs:630000,20.1787918451395)
--(axis cs:620000,20.1915063509461)
--(axis cs:610000,20.2606071321407)
--(axis cs:600000,20.2933034373659)
--(axis cs:590000,20.6172617529929)
--(axis cs:580000,20.2)
--(axis cs:570000,20.143487097856)
--(axis cs:560000,20.1531128874149)
--(axis cs:550000,20.2520797289396)
--(axis cs:540000,20.1847679857416)
--(axis cs:530000,20.2345207879912)
--(axis cs:520000,20.3881459761583)
--(axis cs:510000,20.2949662146737)
--(axis cs:500000,20.791336342194)
--(axis cs:490000,19.1667404100919)
--(axis cs:480000,20.0870828693387)
--(axis cs:470000,20.2701562118716)
--(axis cs:460000,20.161803398875)
--(axis cs:450000,20.5097007781763)
--(axis cs:440000,20.3929603547154)
--(axis cs:430000,20.3652545696244)
--(axis cs:420000,20.5887482193696)
--(axis cs:410000,20.3737468592767)
--(axis cs:400000,20.0338539126016)
--(axis cs:390000,20.1415880433164)
--(axis cs:380000,20.3111380785565)
--(axis cs:370000,20.1123724356958)
--(axis cs:360000,20.2741060010942)
--(axis cs:350000,19.6615773105864)
--(axis cs:340000,20.503965757582)
--(axis cs:330000,19.7077175444876)
--(axis cs:320000,20.1653311931459)
--(axis cs:310000,19.8041522986797)
--(axis cs:300000,20.4844288770225)
--(axis cs:290000,20.1595452979136)
--(axis cs:280000,19.6385164807135)
--(axis cs:270000,19.8007572282457)
--(axis cs:260000,20.0439401922274)
--(axis cs:250000,20.319740215917)
--(axis cs:240000,19.9572904116448)
--(axis cs:230000,19.1123724356958)
--(axis cs:220000,19.4792853971073)
--(axis cs:210000,19.4540218225441)
--(axis cs:200000,17.0094810050209)
--(axis cs:190000,13.3)
--(axis cs:180000,9.80391855782443)
--(axis cs:170000,7.69551939166402)
--(axis cs:160000,5.38415985436581)
--(axis cs:150000,2.04722066167242)
--(axis cs:140000,-0.69364329712605)
--(axis cs:130000,-2.49791847198287)
--(axis cs:120000,-4.16189172389374)
--(axis cs:110000,-6.24304863001465)
--(axis cs:100000,-7.69570239270926)
--(axis cs:90000,-8.19424773770335)
--(axis cs:80000,-8.87812876506571)
--(axis cs:70000,-9.2360125068397)
--(axis cs:60000,-12.6253290043389)
--(axis cs:50000,-14.8901613023126)
--(axis cs:40000,-18.3688048287806)
--(axis cs:30000,-19.4218911086754)
--(axis cs:20000,-19.9388619214435)
--(axis cs:10000,-19.9661460873984)
--(axis cs:0,-17.0642374527743)
--cycle;

\addplot [semithick, steelblue31119180]
table {%
0 -18.1875
10000 -20.5
20000 -20.225
30000 -19.725
40000 -18.725
50000 -15.975
60000 -13.575
70000 -11.075
80000 -10.175
90000 -9.175
100000 -8.425
110000 -7.775
120000 -5.575
130000 -3.7
140000 -1.975
150000 -0.425
160000 2.275
170000 4.35
180000 7.25
190000 12.5
200000 15.95
210000 18.675
220000 18.675
230000 18.5
240000 19.525
250000 19.55
260000 19.325
270000 18.975
280000 19.1
290000 19.5
300000 19.85
310000 19.2
320000 19.75
330000 18.975
340000 17.35
350000 18.9
360000 19.925
370000 19.5
380000 20.025
390000 19.7
400000 19.5
410000 20.125
420000 20.15
430000 19.825
440000 19.775
450000 19.925
460000 20.05
470000 19.95
480000 19.9
490000 18.225
500000 19.425
510000 19.925
520000 20.025
530000 20
540000 19.7
550000 19.65
560000 19.75
570000 19.475
580000 19.85
590000 19.45
600000 19.55
610000 19.725
620000 19.975
630000 19.6
640000 20.2
650000 20.025
660000 19.675
670000 19.975
680000 20.05
690000 19.65
700000 18.375
710000 20.05
720000 20.125
730000 20
740000 20.5
750000 19.925
760000 19.95
770000 19.575
780000 20.025
790000 20.025
800000 19.8
810000 20.025
820000 20.15
830000 20.225
840000 19.975
850000 20.225
860000 19.925
870000 20
880000 19.5
890000 19.3
900000 18.775
910000 19.275
920000 19.225
930000 20.15
940000 20.225
950000 20.275
960000 19.3
970000 20.1
980000 20.175
990000 20.1
1000000 19.925
1010000 20.225
1020000 20.4
1030000 20.25
1040000 19.4
1050000 19.375
1060000 20.2
1070000 19
1080000 19.725
1090000 20.25
};
\addlegendentry{reference (no bottleneck)}

\nextgroupplot[
height=\figheight,
legend cell align={left},
legend style={
  fill opacity=0.8,
  draw opacity=1,
  text opacity=1,
  at={(0.03,0.03)},
  anchor=south west,
  draw=lightgray204
},
tick align=outside,
tick pos=left,
title={Breakout (Classical reference)},
width=\figwidth,
x grid style={darkgray176},
xlabel={Environment steps},
xmajorgrids,
xmin=0, xmax=2500000,
xtick style={color=black},
y grid style={darkgray176},
ymajorgrids,
ymin=-10, ymax=300,
ytick style={color=black}
]
\path [fill=steelblue31119180, fill opacity=0.3]
(axis cs:0,1.02962457555422)
--(axis cs:0,0.450033853428007)
--(axis cs:10000,0.706521584639494)
--(axis cs:20000,0.814412666534069)
--(axis cs:30000,1.08768094211747)
--(axis cs:40000,1.7235601718718)
--(axis cs:50000,2.81345963996323)
--(axis cs:60000,3.95562564202917)
--(axis cs:70000,5.28484387355496)
--(axis cs:80000,6.44772811483255)
--(axis cs:90000,7.77480958004033)
--(axis cs:100000,9.07756412646224)
--(axis cs:110000,10.3947246509951)
--(axis cs:120000,11.9145733140541)
--(axis cs:130000,13.6424118056312)
--(axis cs:140000,15.2942760549426)
--(axis cs:150000,16.9698414359969)
--(axis cs:160000,18.444445205925)
--(axis cs:170000,19.8780426368688)
--(axis cs:180000,20.8748502201391)
--(axis cs:190000,21.501991388164)
--(axis cs:200000,21.3814407087919)
--(axis cs:210000,21.445937522148)
--(axis cs:220000,21.7818851398293)
--(axis cs:230000,22.4984304073457)
--(axis cs:240000,22.5975981785073)
--(axis cs:250000,22.5878605990526)
--(axis cs:260000,22.4046999373813)
--(axis cs:270000,22.8199061985113)
--(axis cs:280000,23.5862318754258)
--(axis cs:290000,24.2842886641609)
--(axis cs:300000,25.0404147697056)
--(axis cs:310000,25.1830819483107)
--(axis cs:320000,26.0542566100113)
--(axis cs:330000,26.8983866465887)
--(axis cs:340000,27.5645240918875)
--(axis cs:350000,27.7387976150523)
--(axis cs:360000,28.1545552046149)
--(axis cs:370000,28.4969774704009)
--(axis cs:380000,29.4499250606741)
--(axis cs:390000,30.6025621888687)
--(axis cs:400000,31.1689473184718)
--(axis cs:410000,32.0592235982404)
--(axis cs:420000,32.7927187617437)
--(axis cs:430000,33.2320970186476)
--(axis cs:440000,33.8965282412919)
--(axis cs:450000,35.4397949924559)
--(axis cs:460000,36.7740042920254)
--(axis cs:470000,36.8332440424643)
--(axis cs:480000,37.9872468245471)
--(axis cs:490000,39.0706796681351)
--(axis cs:500000,41.023934890527)
--(axis cs:510000,41.899073663745)
--(axis cs:520000,44.160398828475)
--(axis cs:530000,44.9762635742002)
--(axis cs:540000,48.7969892830746)
--(axis cs:550000,51.7424043925643)
--(axis cs:560000,54.5387558982494)
--(axis cs:570000,57.8375443877504)
--(axis cs:580000,58.9984408793465)
--(axis cs:590000,59.7380803647861)
--(axis cs:600000,61.4042882289209)
--(axis cs:610000,62.3025561420798)
--(axis cs:620000,63.7679623214929)
--(axis cs:630000,66.3116008066106)
--(axis cs:640000,66.0679083685056)
--(axis cs:650000,67.2131560604308)
--(axis cs:660000,66.9189729682048)
--(axis cs:670000,67.5510792590099)
--(axis cs:680000,68.4804651621956)
--(axis cs:690000,69.3194783587956)
--(axis cs:700000,70.356332905521)
--(axis cs:710000,72.3177715640452)
--(axis cs:720000,74.9057112071355)
--(axis cs:730000,77.1957657990928)
--(axis cs:740000,77.8602726396975)
--(axis cs:750000,81.7065030062794)
--(axis cs:760000,82.4180229918885)
--(axis cs:770000,85.7277553949441)
--(axis cs:780000,88.1136373843347)
--(axis cs:790000,89.7580769979781)
--(axis cs:800000,94.170393495254)
--(axis cs:810000,97.9565408057347)
--(axis cs:820000,103.328566585371)
--(axis cs:830000,105.343924389449)
--(axis cs:840000,107.897651596519)
--(axis cs:850000,112.232412519347)
--(axis cs:860000,119.186456888066)
--(axis cs:870000,121.233364468446)
--(axis cs:880000,122.018167089416)
--(axis cs:890000,125.513409010783)
--(axis cs:900000,128.575117425244)
--(axis cs:910000,131.380330805279)
--(axis cs:920000,134.697353213475)
--(axis cs:930000,136.418694743837)
--(axis cs:940000,140.442771477576)
--(axis cs:950000,143.113538450519)
--(axis cs:960000,145.92185601901)
--(axis cs:970000,144.113339455871)
--(axis cs:980000,146.455630345482)
--(axis cs:990000,149.526245103154)
--(axis cs:1000000,148.445119350902)
--(axis cs:1010000,150.779118949287)
--(axis cs:1020000,149.269742534047)
--(axis cs:1030000,145.3409942209)
--(axis cs:1040000,147.452310124279)
--(axis cs:1050000,152.241334464154)
--(axis cs:1060000,148.545025720434)
--(axis cs:1070000,150.004688574636)
--(axis cs:1080000,148.219987187944)
--(axis cs:1090000,152.96686165421)
--(axis cs:1100000,155.35276912468)
--(axis cs:1110000,158.541186353707)
--(axis cs:1120000,161.374369376415)
--(axis cs:1130000,162.882992122782)
--(axis cs:1140000,166.079408194217)
--(axis cs:1150000,166.762097198692)
--(axis cs:1160000,168.952070991715)
--(axis cs:1170000,173.604484331124)
--(axis cs:1180000,179.209964396852)
--(axis cs:1190000,181.294748731101)
--(axis cs:1200000,183.462556787643)
--(axis cs:1210000,186.401043591312)
--(axis cs:1220000,188.729460086798)
--(axis cs:1230000,189.664519101833)
--(axis cs:1240000,191.848211313626)
--(axis cs:1250000,193.710957387276)
--(axis cs:1260000,194.236145743846)
--(axis cs:1270000,194.341532696012)
--(axis cs:1280000,195.014237893622)
--(axis cs:1290000,197.848908226943)
--(axis cs:1300000,198.184479530022)
--(axis cs:1310000,199.934908622008)
--(axis cs:1320000,199.954610250997)
--(axis cs:1330000,201.099590033384)
--(axis cs:1340000,201.333380489399)
--(axis cs:1350000,203.241785519975)
--(axis cs:1360000,203.061094426811)
--(axis cs:1370000,205.20791743512)
--(axis cs:1380000,204.790269096244)
--(axis cs:1390000,208.363031949417)
--(axis cs:1400000,209.795080482338)
--(axis cs:1410000,208.708865097847)
--(axis cs:1420000,208.84955670754)
--(axis cs:1430000,209.704867512786)
--(axis cs:1440000,207.915314668552)
--(axis cs:1450000,209.168319024995)
--(axis cs:1460000,212.040348164809)
--(axis cs:1470000,215.858363177421)
--(axis cs:1480000,220.663526177542)
--(axis cs:1490000,220.208171750413)
--(axis cs:1500000,218.68153702603)
--(axis cs:1510000,216.205073385622)
--(axis cs:1520000,218.984242782093)
--(axis cs:1530000,221.244071849628)
--(axis cs:1540000,226.468397574778)
--(axis cs:1550000,229.192312174117)
--(axis cs:1560000,224.237625385211)
--(axis cs:1570000,231.139867845055)
--(axis cs:1580000,230.597476979614)
--(axis cs:1590000,234.045908973446)
--(axis cs:1600000,231.537995057017)
--(axis cs:1610000,234.096399393109)
--(axis cs:1620000,234.826137618842)
--(axis cs:1630000,237.468241104007)
--(axis cs:1640000,238.750467635553)
--(axis cs:1650000,236.423626002749)
--(axis cs:1660000,239.579364548903)
--(axis cs:1670000,240.201711604093)
--(axis cs:1680000,239.208379534038)
--(axis cs:1690000,234.319823397562)
--(axis cs:1700000,233.174486444147)
--(axis cs:1710000,230.478626343985)
--(axis cs:1720000,227.058929996906)
--(axis cs:1730000,231.291859490814)
--(axis cs:1740000,227.888658129343)
--(axis cs:1750000,232.165079856175)
--(axis cs:1760000,231.605950642418)
--(axis cs:1770000,228.030221370422)
--(axis cs:1780000,232.056241467883)
--(axis cs:1790000,231.859709315733)
--(axis cs:1800000,233.415857072956)
--(axis cs:1810000,235.551396220369)
--(axis cs:1820000,236.535409852851)
--(axis cs:1830000,238.60524472899)
--(axis cs:1840000,242.726158048309)
--(axis cs:1850000,246.379909418391)
--(axis cs:1860000,246.900988943314)
--(axis cs:1870000,242.11427996229)
--(axis cs:1880000,242.028492090743)
--(axis cs:1890000,244.024293446827)
--(axis cs:1900000,245.274926947351)
--(axis cs:1910000,248.652620877343)
--(axis cs:1920000,252.834559267034)
--(axis cs:1930000,250.830149991475)
--(axis cs:1940000,251.00026690692)
--(axis cs:1950000,252.041408132454)
--(axis cs:1960000,245.145596003657)
--(axis cs:1970000,247.194168637551)
--(axis cs:1980000,249.438215438243)
--(axis cs:1990000,250.666377668814)
--(axis cs:2000000,250.531450450375)
--(axis cs:2010000,252.835669916058)
--(axis cs:2020000,245.921224328123)
--(axis cs:2030000,248.280059830345)
--(axis cs:2040000,248.819540631993)
--(axis cs:2050000,249.789911445744)
--(axis cs:2060000,250.425945861896)
--(axis cs:2070000,250.350656877222)
--(axis cs:2080000,247.183904967705)
--(axis cs:2090000,245.024694779224)
--(axis cs:2100000,243.503432614996)
--(axis cs:2110000,238.732817066751)
--(axis cs:2120000,238.487809166526)
--(axis cs:2130000,237.608123845872)
--(axis cs:2140000,233.242420117466)
--(axis cs:2150000,236.208501902462)
--(axis cs:2160000,238.87486252063)
--(axis cs:2170000,234.894669190985)
--(axis cs:2180000,236.057510534136)
--(axis cs:2190000,238.626771699846)
--(axis cs:2200000,245.321323928689)
--(axis cs:2210000,243.887629895188)
--(axis cs:2220000,245.448427454413)
--(axis cs:2230000,243.874453351931)
--(axis cs:2240000,248.028820030761)
--(axis cs:2250000,244.323712994934)
--(axis cs:2260000,240.024669240401)
--(axis cs:2270000,235.707024030503)
--(axis cs:2280000,234.315163520993)
--(axis cs:2290000,233.993718778292)
--(axis cs:2300000,234.025636181537)
--(axis cs:2310000,236.413129675265)
--(axis cs:2320000,237.113430894253)
--(axis cs:2330000,242.033230544343)
--(axis cs:2340000,245.61623322597)
--(axis cs:2350000,243.336547330591)
--(axis cs:2360000,245.46621815141)
--(axis cs:2370000,238.118753953859)
--(axis cs:2380000,239.388471499501)
--(axis cs:2390000,232.30971968392)
--(axis cs:2400000,226.671328377256)
--(axis cs:2410000,225.887393656362)
--(axis cs:2420000,223.461633041875)
--(axis cs:2430000,219.976527851898)
--(axis cs:2440000,222.267975506115)
--(axis cs:2450000,228.396290225208)
--(axis cs:2460000,231.78349350832)
--(axis cs:2470000,234.522721152393)
--(axis cs:2480000,237.638043851019)
--(axis cs:2490000,237.781725273988)
--(axis cs:2490000,297.549474726012)
--(axis cs:2490000,297.549474726012)
--(axis cs:2480000,297.101156148981)
--(axis cs:2470000,292.702878847607)
--(axis cs:2460000,297.133306491679)
--(axis cs:2450000,295.907709774792)
--(axis cs:2440000,298.773624493885)
--(axis cs:2430000,309.361072148102)
--(axis cs:2420000,307.815166958125)
--(axis cs:2410000,301.951006343638)
--(axis cs:2400000,296.744671622744)
--(axis cs:2390000,296.04548031608)
--(axis cs:2380000,297.304328500499)
--(axis cs:2370000,295.046046046141)
--(axis cs:2360000,294.12578184859)
--(axis cs:2350000,298.636252669409)
--(axis cs:2340000,297.26856677403)
--(axis cs:2330000,290.901169455657)
--(axis cs:2320000,283.808169105747)
--(axis cs:2310000,272.686070324735)
--(axis cs:2300000,266.607963818463)
--(axis cs:2290000,264.889481221708)
--(axis cs:2280000,266.579236479007)
--(axis cs:2270000,265.257775969497)
--(axis cs:2260000,262.245730759599)
--(axis cs:2250000,261.543487005066)
--(axis cs:2240000,258.44477996924)
--(axis cs:2230000,255.498346648069)
--(axis cs:2220000,258.461972545587)
--(axis cs:2210000,269.753970104812)
--(axis cs:2200000,271.536276071311)
--(axis cs:2190000,271.322028300154)
--(axis cs:2180000,271.539289465864)
--(axis cs:2170000,272.361330809015)
--(axis cs:2160000,267.66273747937)
--(axis cs:2150000,268.466698097538)
--(axis cs:2140000,269.477579882534)
--(axis cs:2130000,266.118276154128)
--(axis cs:2120000,263.976190833474)
--(axis cs:2110000,262.395182933249)
--(axis cs:2100000,266.443767385004)
--(axis cs:2090000,264.360905220776)
--(axis cs:2080000,267.960095032295)
--(axis cs:2070000,268.343743122778)
--(axis cs:2060000,273.286054138105)
--(axis cs:2050000,283.654888554256)
--(axis cs:2040000,288.294059368008)
--(axis cs:2030000,289.377540169655)
--(axis cs:2020000,286.493175671877)
--(axis cs:2010000,291.375530083942)
--(axis cs:2000000,297.742149549625)
--(axis cs:1990000,301.565622331186)
--(axis cs:1980000,302.400184561757)
--(axis cs:1970000,294.740231362449)
--(axis cs:1960000,294.150403996343)
--(axis cs:1950000,293.873791867546)
--(axis cs:1940000,289.47813309308)
--(axis cs:1930000,290.419450008525)
--(axis cs:1920000,283.535040732966)
--(axis cs:1910000,281.041779122657)
--(axis cs:1900000,277.501073052649)
--(axis cs:1890000,276.361306553173)
--(axis cs:1880000,288.565107909257)
--(axis cs:1870000,287.55612003771)
--(axis cs:1860000,287.734211056686)
--(axis cs:1850000,285.402490581609)
--(axis cs:1840000,290.704241951691)
--(axis cs:1830000,292.78035527101)
--(axis cs:1820000,288.536590147149)
--(axis cs:1810000,287.101403779631)
--(axis cs:1800000,287.404942927044)
--(axis cs:1790000,288.393090684266)
--(axis cs:1780000,288.775758532117)
--(axis cs:1770000,291.356978629578)
--(axis cs:1760000,291.197249357582)
--(axis cs:1750000,290.802920143825)
--(axis cs:1740000,288.708141870657)
--(axis cs:1730000,288.191340509186)
--(axis cs:1720000,286.246670003094)
--(axis cs:1710000,285.457373656015)
--(axis cs:1700000,286.924713555853)
--(axis cs:1690000,286.620976602438)
--(axis cs:1680000,290.610820465962)
--(axis cs:1670000,289.054288395907)
--(axis cs:1660000,284.972635451097)
--(axis cs:1650000,286.147573997251)
--(axis cs:1640000,286.204732364447)
--(axis cs:1630000,284.592558895993)
--(axis cs:1620000,281.685862381158)
--(axis cs:1610000,280.233200606891)
--(axis cs:1600000,282.260404942983)
--(axis cs:1590000,280.040491026554)
--(axis cs:1580000,277.887323020386)
--(axis cs:1570000,274.604132154945)
--(axis cs:1560000,274.912774614789)
--(axis cs:1550000,277.314887825883)
--(axis cs:1540000,274.958802425222)
--(axis cs:1530000,271.687128150372)
--(axis cs:1520000,273.018957217907)
--(axis cs:1510000,269.652526614378)
--(axis cs:1500000,271.08166297397)
--(axis cs:1490000,270.297428249587)
--(axis cs:1480000,272.522073822458)
--(axis cs:1470000,271.128836822578)
--(axis cs:1460000,269.820451835191)
--(axis cs:1450000,271.684480975005)
--(axis cs:1440000,266.937485331448)
--(axis cs:1430000,266.061532487214)
--(axis cs:1420000,265.39524329246)
--(axis cs:1410000,265.003134902153)
--(axis cs:1400000,267.145719517662)
--(axis cs:1390000,264.993768050583)
--(axis cs:1380000,269.072130903756)
--(axis cs:1370000,273.47528256488)
--(axis cs:1360000,272.750105573189)
--(axis cs:1350000,271.526214480025)
--(axis cs:1340000,270.741819510601)
--(axis cs:1330000,267.481209966616)
--(axis cs:1320000,267.573389749003)
--(axis cs:1310000,268.495491377992)
--(axis cs:1300000,269.050720469978)
--(axis cs:1290000,272.181491773057)
--(axis cs:1280000,271.144162106378)
--(axis cs:1270000,270.319267303988)
--(axis cs:1260000,270.731854256154)
--(axis cs:1250000,267.175442612724)
--(axis cs:1240000,265.193388686374)
--(axis cs:1230000,265.465080898168)
--(axis cs:1220000,263.427339913202)
--(axis cs:1210000,263.474156408688)
--(axis cs:1200000,264.342243212357)
--(axis cs:1190000,262.906851268899)
--(axis cs:1180000,263.081235603148)
--(axis cs:1170000,261.017915668876)
--(axis cs:1160000,258.724729008285)
--(axis cs:1150000,253.717902801308)
--(axis cs:1140000,247.234191805783)
--(axis cs:1130000,245.384207877218)
--(axis cs:1120000,240.476830623585)
--(axis cs:1110000,242.385213646293)
--(axis cs:1100000,239.80083087532)
--(axis cs:1090000,236.46993834579)
--(axis cs:1080000,229.743212812056)
--(axis cs:1070000,226.563311425364)
--(axis cs:1060000,225.235774279566)
--(axis cs:1050000,224.369865535846)
--(axis cs:1040000,218.261289875721)
--(axis cs:1030000,214.1614057791)
--(axis cs:1020000,208.490257465953)
--(axis cs:1010000,205.105681050713)
--(axis cs:1000000,202.516480649098)
--(axis cs:990000,204.512154896846)
--(axis cs:980000,200.157169654518)
--(axis cs:970000,197.505860544129)
--(axis cs:960000,197.84774398099)
--(axis cs:950000,191.912061549481)
--(axis cs:940000,190.834028522424)
--(axis cs:930000,187.358905256163)
--(axis cs:920000,180.822646786525)
--(axis cs:910000,179.141269194721)
--(axis cs:900000,175.303282574756)
--(axis cs:890000,172.075390989217)
--(axis cs:880000,169.077832910584)
--(axis cs:870000,166.505835531554)
--(axis cs:860000,163.669543111934)
--(axis cs:850000,159.276387480653)
--(axis cs:840000,153.097548403481)
--(axis cs:830000,152.177675610551)
--(axis cs:820000,147.925833414629)
--(axis cs:810000,148.955459194265)
--(axis cs:800000,147.596006504746)
--(axis cs:790000,143.405123002022)
--(axis cs:780000,141.012762615665)
--(axis cs:770000,141.806644605056)
--(axis cs:760000,137.484377008111)
--(axis cs:750000,138.178296993721)
--(axis cs:740000,136.803727360302)
--(axis cs:730000,134.549834200907)
--(axis cs:720000,132.427088792864)
--(axis cs:710000,129.224628435955)
--(axis cs:700000,129.437267094479)
--(axis cs:690000,126.088521641204)
--(axis cs:680000,125.055534837804)
--(axis cs:670000,123.15452074099)
--(axis cs:660000,121.407427031795)
--(axis cs:650000,118.450843939569)
--(axis cs:640000,115.588091631494)
--(axis cs:630000,110.942799193389)
--(axis cs:620000,108.016037678507)
--(axis cs:610000,105.62704385792)
--(axis cs:600000,104.485311771079)
--(axis cs:590000,100.020319635214)
--(axis cs:580000,97.5743591206535)
--(axis cs:570000,94.1576556122497)
--(axis cs:560000,91.2212441017506)
--(axis cs:550000,88.7503956074357)
--(axis cs:540000,85.0126107169254)
--(axis cs:530000,81.9981364257997)
--(axis cs:520000,78.986801171525)
--(axis cs:510000,75.683326336255)
--(axis cs:500000,72.262465109473)
--(axis cs:490000,68.7261203318649)
--(axis cs:480000,65.2959531754529)
--(axis cs:470000,62.8771559575357)
--(axis cs:460000,59.7747957079746)
--(axis cs:450000,58.5010050075441)
--(axis cs:440000,54.9978717587081)
--(axis cs:430000,52.4415029813524)
--(axis cs:420000,49.5336812382563)
--(axis cs:410000,47.3823764017596)
--(axis cs:400000,45.5702526815283)
--(axis cs:390000,43.2454378111313)
--(axis cs:380000,41.0796749393259)
--(axis cs:370000,39.7302225295991)
--(axis cs:360000,37.8118447953851)
--(axis cs:350000,36.7956023849477)
--(axis cs:340000,35.4978759081125)
--(axis cs:330000,34.3272133534113)
--(axis cs:320000,33.2225433899887)
--(axis cs:310000,31.6617180516893)
--(axis cs:300000,30.6843852302944)
--(axis cs:290000,29.7669113358391)
--(axis cs:280000,28.9081681245742)
--(axis cs:270000,28.0872938014887)
--(axis cs:260000,27.3553000626187)
--(axis cs:250000,26.7337394009474)
--(axis cs:240000,26.1112018214927)
--(axis cs:230000,25.7607695926543)
--(axis cs:220000,24.8229148601707)
--(axis cs:210000,24.782862477852)
--(axis cs:200000,24.6681592912081)
--(axis cs:190000,24.158808611836)
--(axis cs:180000,23.2483497798609)
--(axis cs:170000,21.8931573631313)
--(axis cs:160000,20.571554794075)
--(axis cs:150000,18.9709585640031)
--(axis cs:140000,17.2049239450574)
--(axis cs:130000,15.4855881943688)
--(axis cs:120000,13.6758266859459)
--(axis cs:110000,12.0404753490049)
--(axis cs:100000,10.5048358735378)
--(axis cs:90000,8.86039041995967)
--(axis cs:80000,7.53787188516745)
--(axis cs:70000,6.02875612644504)
--(axis cs:60000,4.61077435797083)
--(axis cs:50000,3.11454036003677)
--(axis cs:40000,2.1052398281282)
--(axis cs:30000,1.42112476568103)
--(axis cs:20000,1.27671784129325)
--(axis cs:10000,1.21413700565922)
--(axis cs:0,1.02962457555422)
--cycle;

\addplot [semithick, steelblue31119180]
table {%
0 0.739829214491115
10000 0.960329295149357
20000 1.04556525391366
30000 1.25440285389925
40000 1.9144
50000 2.964
60000 4.2832
70000 5.6568
80000 6.9928
90000 8.3176
100000 9.7912
110000 11.2176
120000 12.7952
130000 14.564
140000 16.2496
150000 17.9704
160000 19.508
170000 20.8856
180000 22.0616
190000 22.8304
200000 23.0248
210000 23.1144
220000 23.3024
230000 24.1296
240000 24.3544
250000 24.6608
260000 24.88
270000 25.4536
280000 26.2472
290000 27.0256
300000 27.8624
310000 28.4224
320000 29.6384
330000 30.6128
340000 31.5312
350000 32.2672
360000 32.9832
370000 34.1136
380000 35.2648
390000 36.924
400000 38.3696
410000 39.7208
420000 41.1632
430000 42.8368
440000 44.4472
450000 46.9704
460000 48.2744
470000 49.8552
480000 51.6416
490000 53.8984
500000 56.6432
510000 58.7912
520000 61.5736
530000 63.4872
540000 66.9048
550000 70.2464
560000 72.88
570000 75.9976
580000 78.2864
590000 79.8792
600000 82.9448
610000 83.9648
620000 85.892
630000 88.6272
640000 90.828
650000 92.832
660000 94.1632
670000 95.3528
680000 96.768
690000 97.704
700000 99.8968
710000 100.7712
720000 103.6664
730000 105.8728
740000 107.332
750000 109.9424
760000 109.9512
770000 113.7672
780000 114.5632
790000 116.5816
800000 120.8832
810000 123.456
820000 125.6272
830000 128.7608
840000 130.4976
850000 135.7544
860000 141.428
870000 143.8696
880000 145.548
890000 148.7944
900000 151.9392
910000 155.2608
920000 157.76
930000 161.8888
940000 165.6384
950000 167.5128
960000 171.8848
970000 170.8096
980000 173.3064
990000 177.0192
1000000 175.4808
1010000 177.9424
1020000 178.88
1030000 179.7512
1040000 182.8568
1050000 188.3056
1060000 186.8904
1070000 188.284
1080000 188.9816
1090000 194.7184
1100000 197.5768
1110000 200.4632
1120000 200.9256
1130000 204.1336
1140000 206.6568
1150000 210.24
1160000 213.8384
1170000 217.3112
1180000 221.1456
1190000 222.1008
1200000 223.9024
1210000 224.9376
1220000 226.0784
1230000 227.5648
1240000 228.5208
1250000 230.4432
1260000 232.484
1270000 232.3304
1280000 233.0792
1290000 235.0152
1300000 233.6176
1310000 234.2152
1320000 233.764
1330000 234.2904
1340000 236.0376
1350000 237.384
1360000 237.9056
1370000 239.3416
1380000 236.9312
1390000 236.6784
1400000 238.4704
1410000 236.856
1420000 237.1224
1430000 237.8832
1440000 237.4264
1450000 240.4264
1460000 240.9304
1470000 243.4936
1480000 246.5928
1490000 245.2528
1500000 244.8816
1510000 242.9288
1520000 246.0016
1530000 246.4656
1540000 250.7136
1550000 253.2536
1560000 249.5752
1570000 252.872
1580000 254.2424
1590000 257.0432
1600000 256.8992
1610000 257.1648
1620000 258.256
1630000 261.0304
1640000 262.4776
1650000 261.2856
1660000 262.276
1670000 264.628
1680000 264.9096
1690000 260.4704
1700000 260.0496
1710000 257.968
1720000 256.6528
1730000 259.7416
1740000 258.2984
1750000 261.484
1760000 261.4016
1770000 259.6936
1780000 260.416
1790000 260.1264
1800000 260.4104
1810000 261.3264
1820000 262.536
1830000 265.6928
1840000 266.7152
1850000 265.8912
1860000 267.3176
1870000 264.8352
1880000 265.2968
1890000 260.1928
1900000 261.388
1910000 264.8472
1920000 268.1848
1930000 270.6248
1940000 270.2392
1950000 272.9576
1960000 269.648
1970000 270.9672
1980000 275.9192
1990000 276.116
2000000 274.1368
2010000 272.1056
2020000 266.2072
2030000 268.8288
2040000 268.5568
2050000 266.7224
2060000 261.856
2070000 259.3472
2080000 257.572
2090000 254.6928
2100000 254.9736
2110000 250.564
2120000 251.232
2130000 251.8632
2140000 251.36
2150000 252.3376
2160000 253.2688
2170000 253.628
2180000 253.7984
2190000 254.9744
2200000 258.4288
2210000 256.8208
2220000 251.9552
2230000 249.6864
2240000 253.2368
2250000 252.9336
2260000 251.1352
2270000 250.4824
2280000 250.4472
2290000 249.4416
2300000 250.3168
2310000 254.5496
2320000 260.4608
2330000 266.4672
2340000 271.4424
2350000 270.9864
2360000 269.796
2370000 266.5824
2380000 268.3464
2390000 264.1776
2400000 261.708
2410000 263.9192
2420000 265.6384
2430000 264.6688
2440000 260.5208
2450000 262.152
2460000 264.4584
2470000 263.6128
2480000 267.3696
2490000 267.6656
};
\addlegendentry{reference (no bottleneck)}
\end{groupplot}

\end{tikzpicture}

%% file: graphics/figures/feature-distributions.tex
\begin{tikzpicture}

\definecolor{darkgray176}{RGB}{176,176,176}
\definecolor{darkslategray66}{RGB}{66,66,66}
\definecolor{dimgray10784138}{RGB}{107,84,138}
\definecolor{gray}{RGB}{128,128,128}
\definecolor{salmon234109113}{RGB}{234,109,113}
\definecolor{steelblue46125174}{RGB}{46,125,174}

\begin{groupplot}[group style={group size=2 by 1}]
\nextgroupplot[
height=\figheight,
tick align=outside,
tick pos=left,
title={Feature values (Hybrid model)},
width=\figwidth,
x grid style={darkgray176},
xlabel={Latent feature},
xmin=-0.5, xmax=15.5,
xtick style={color=black},
y grid style={darkgray176},
ylabel={Value},
ymajorgrids,
ymin=-3.61160362958908, ymax=3.4578496336937,
ytick style={color=black},
ytick={-3, -2, -1, 0, 1, 2, 3}
]
\path [draw=darkslategray66, fill=steelblue46125174, semithick]
(axis cs:-0.4,0.0437655635178089)
--(axis cs:0.4,0.0437655635178089)
--(axis cs:0.4,0.945286944508553)
--(axis cs:-0.4,0.945286944508553)
--(axis cs:-0.4,0.0437655635178089)
--cycle;
\path [draw=darkslategray66, fill=dimgray10784138, semithick]
(axis cs:0.6,-0.292865365743637)
--(axis cs:1.4,-0.292865365743637)
--(axis cs:1.4,0.118476040661335)
--(axis cs:0.6,0.118476040661335)
--(axis cs:0.6,-0.292865365743637)
--cycle;
\path [draw=darkslategray66, fill=salmon234109113, semithick]
(axis cs:1.6,-1.92662909626961)
--(axis cs:2.4,-1.92662909626961)
--(axis cs:2.4,-0.986063629388809)
--(axis cs:1.6,-0.986063629388809)
--(axis cs:1.6,-1.92662909626961)
--cycle;
\path [draw=darkslategray66, fill=gray, semithick]
(axis cs:2.6,0.484660781919956)
--(axis cs:3.4,0.484660781919956)
--(axis cs:3.4,1.93989503383636)
--(axis cs:2.6,1.93989503383636)
--(axis cs:2.6,0.484660781919956)
--cycle;
\path [draw=darkslategray66, fill=steelblue46125174, semithick]
(axis cs:3.6,0.194086633622646)
--(axis cs:4.4,0.194086633622646)
--(axis cs:4.4,1.12191939353943)
--(axis cs:3.6,1.12191939353943)
--(axis cs:3.6,0.194086633622646)
--cycle;
\path [draw=darkslategray66, fill=dimgray10784138, semithick]
(axis cs:4.6,-0.495936259627342)
--(axis cs:5.4,-0.495936259627342)
--(axis cs:5.4,0.309945896267891)
--(axis cs:4.6,0.309945896267891)
--(axis cs:4.6,-0.495936259627342)
--cycle;
\path [draw=darkslategray66, fill=salmon234109113, semithick]
(axis cs:5.6,-1.18780663609505)
--(axis cs:6.4,-1.18780663609505)
--(axis cs:6.4,-0.528086349368095)
--(axis cs:5.6,-0.528086349368095)
--(axis cs:5.6,-1.18780663609505)
--cycle;
\path [draw=darkslategray66, fill=gray, semithick]
(axis cs:6.6,0.159584749490023)
--(axis cs:7.4,0.159584749490023)
--(axis cs:7.4,1.14234936237335)
--(axis cs:6.6,1.14234936237335)
--(axis cs:6.6,0.159584749490023)
--cycle;
\path [draw=darkslategray66, fill=steelblue46125174, semithick]
(axis cs:7.6,-0.860394775867462)
--(axis cs:8.4,-0.860394775867462)
--(axis cs:8.4,-0.295139037072659)
--(axis cs:7.6,-0.295139037072659)
--(axis cs:7.6,-0.860394775867462)
--cycle;
\path [draw=darkslategray66, fill=dimgray10784138, semithick]
(axis cs:8.6,-0.831656217575073)
--(axis cs:9.4,-0.831656217575073)
--(axis cs:9.4,-0.0140068084001541)
--(axis cs:8.6,-0.0140068084001541)
--(axis cs:8.6,-0.831656217575073)
--cycle;
\path [draw=darkslategray66, fill=salmon234109113, semithick]
(axis cs:9.6,-0.226614471524954)
--(axis cs:10.4,-0.226614471524954)
--(axis cs:10.4,1.13391208648682)
--(axis cs:9.6,1.13391208648682)
--(axis cs:9.6,-0.226614471524954)
--cycle;
\path [draw=darkslategray66, fill=gray, semithick]
(axis cs:10.6,0.148122116923332)
--(axis cs:11.4,0.148122116923332)
--(axis cs:11.4,1.53612905740738)
--(axis cs:10.6,1.53612905740738)
--(axis cs:10.6,0.148122116923332)
--cycle;
\path [draw=darkslategray66, fill=steelblue46125174, semithick]
(axis cs:11.6,-0.648081570863724)
--(axis cs:12.4,-0.648081570863724)
--(axis cs:12.4,-0.153626337647438)
--(axis cs:11.6,-0.153626337647438)
--(axis cs:11.6,-0.648081570863724)
--cycle;
\path [draw=darkslategray66, fill=dimgray10784138, semithick]
(axis cs:12.6,-0.32913251966238)
--(axis cs:13.4,-0.32913251966238)
--(axis cs:13.4,0.278933241963387)
--(axis cs:12.6,0.278933241963387)
--(axis cs:12.6,-0.32913251966238)
--cycle;
\path [draw=darkslategray66, fill=salmon234109113, semithick]
(axis cs:13.6,-0.381361827254295)
--(axis cs:14.4,-0.381361827254295)
--(axis cs:14.4,0.493440732359886)
--(axis cs:13.6,0.493440732359886)
--(axis cs:13.6,-0.381361827254295)
--cycle;
\path [draw=darkslategray66, fill=gray, semithick]
(axis cs:14.6,0.140507146716118)
--(axis cs:15.4,0.140507146716118)
--(axis cs:15.4,0.910328894853592)
--(axis cs:14.6,0.910328894853592)
--(axis cs:14.6,0.140507146716118)
--cycle;
\addplot [semithick, darkslategray66]
table {%
0 0.0437655635178089
0 -0.789626181125641
};
\addplot [semithick, darkslategray66]
table {%
0 0.945286944508553
0 1.84576261043549
};
\addplot [semithick, darkslategray66]
table {%
-0.2 -0.789626181125641
0.2 -0.789626181125641
};
\addplot [semithick, darkslategray66]
table {%
-0.2 1.84576261043549
0.2 1.84576261043549
};
\addplot [semithick, darkslategray66]
table {%
1 -0.292865365743637
1 -0.907232403755188
};
\addplot [semithick, darkslategray66]
table {%
1 0.118476040661335
1 0.588002324104309
};
\addplot [semithick, darkslategray66]
table {%
0.8 -0.907232403755188
1.2 -0.907232403755188
};
\addplot [semithick, darkslategray66]
table {%
0.8 0.588002324104309
1.2 0.588002324104309
};
\addplot [black, mark=diamond*, mark size=2.5, mark options={solid,fill=darkslategray66}, only marks]
table {%
1 -1.0728542804718
1 -0.989771366119385
1 -1.11937475204468
1 -1.30282330513
1 -1.0352109670639
1 -1.00770604610443
1 0.815306901931763
};
\addplot [semithick, darkslategray66]
table {%
2 -1.92662909626961
2 -3.29026484489441
};
\addplot [semithick, darkslategray66]
table {%
2 -0.986063629388809
2 0.35661906003952
};
\addplot [semithick, darkslategray66]
table {%
1.8 -3.29026484489441
2.2 -3.29026484489441
};
\addplot [semithick, darkslategray66]
table {%
1.8 0.35661906003952
2.2 0.35661906003952
};
\addplot [semithick, darkslategray66]
table {%
3 0.484660781919956
3 -0.668518662452698
};
\addplot [semithick, darkslategray66]
table {%
3 1.93989503383636
3 2.78684854507446
};
\addplot [semithick, darkslategray66]
table {%
2.8 -0.668518662452698
3.2 -0.668518662452698
};
\addplot [semithick, darkslategray66]
table {%
2.8 2.78684854507446
3.2 2.78684854507446
};
\addplot [semithick, darkslategray66]
table {%
4 0.194086633622646
4 -0.597187936306
};
\addplot [semithick, darkslategray66]
table {%
4 1.12191939353943
4 1.63213646411896
};
\addplot [semithick, darkslategray66]
table {%
3.8 -0.597187936306
4.2 -0.597187936306
};
\addplot [semithick, darkslategray66]
table {%
3.8 1.63213646411896
4.2 1.63213646411896
};
\addplot [semithick, darkslategray66]
table {%
5 -0.495936259627342
5 -1.37173509597778
};
\addplot [semithick, darkslategray66]
table {%
5 0.309945896267891
5 1.4717184305191
};
\addplot [semithick, darkslategray66]
table {%
4.8 -1.37173509597778
5.2 -1.37173509597778
};
\addplot [semithick, darkslategray66]
table {%
4.8 1.4717184305191
5.2 1.4717184305191
};
\addplot [black, mark=diamond*, mark size=2.5, mark options={solid,fill=darkslategray66}, only marks]
table {%
5 1.57514750957489
5 1.92617964744568
5 1.79577422142029
};
\addplot [semithick, darkslategray66]
table {%
6 -1.18780663609505
6 -1.91563308238983
};
\addplot [semithick, darkslategray66]
table {%
6 -0.528086349368095
6 0.367659360170364
};
\addplot [semithick, darkslategray66]
table {%
5.8 -1.91563308238983
6.2 -1.91563308238983
};
\addplot [semithick, darkslategray66]
table {%
5.8 0.367659360170364
6.2 0.367659360170364
};
\addplot [semithick, darkslategray66]
table {%
7 0.159584749490023
7 -1.24594330787659
};
\addplot [semithick, darkslategray66]
table {%
7 1.14234936237335
7 2.17122006416321
};
\addplot [semithick, darkslategray66]
table {%
6.8 -1.24594330787659
7.2 -1.24594330787659
};
\addplot [semithick, darkslategray66]
table {%
6.8 2.17122006416321
7.2 2.17122006416321
};
\addplot [semithick, darkslategray66]
table {%
8 -0.860394775867462
8 -1.35388934612274
};
\addplot [semithick, darkslategray66]
table {%
8 -0.295139037072659
8 0.514946937561035
};
\addplot [semithick, darkslategray66]
table {%
7.8 -1.35388934612274
8.2 -1.35388934612274
};
\addplot [semithick, darkslategray66]
table {%
7.8 0.514946937561035
8.2 0.514946937561035
};
\addplot [black, mark=diamond*, mark size=2.5, mark options={solid,fill=darkslategray66}, only marks]
table {%
8 0.57611620426178
8 0.638243794441223
8 1.02586579322815
};
\addplot [semithick, darkslategray66]
table {%
9 -0.831656217575073
9 -1.53385245800018
};
\addplot [semithick, darkslategray66]
table {%
9 -0.0140068084001541
9 0.933429956436157
};
\addplot [semithick, darkslategray66]
table {%
8.8 -1.53385245800018
9.2 -1.53385245800018
};
\addplot [semithick, darkslategray66]
table {%
8.8 0.933429956436157
9.2 0.933429956436157
};
\addplot [black, mark=diamond*, mark size=2.5, mark options={solid,fill=darkslategray66}, only marks]
table {%
9 1.40009498596191
};
\addplot [semithick, darkslategray66]
table {%
10 -0.226614471524954
10 -1.56565976142883
};
\addplot [semithick, darkslategray66]
table {%
10 1.13391208648682
10 2.42344856262207
};
\addplot [semithick, darkslategray66]
table {%
9.8 -1.56565976142883
10.2 -1.56565976142883
};
\addplot [semithick, darkslategray66]
table {%
9.8 2.42344856262207
10.2 2.42344856262207
};
\addplot [semithick, darkslategray66]
table {%
11 0.148122116923332
11 -1.75598466396332
};
\addplot [semithick, darkslategray66]
table {%
11 1.53612905740738
11 3.13651084899902
};
\addplot [semithick, darkslategray66]
table {%
10.8 -1.75598466396332
11.2 -1.75598466396332
};
\addplot [semithick, darkslategray66]
table {%
10.8 3.13651084899902
11.2 3.13651084899902
};
\addplot [semithick, darkslategray66]
table {%
12 -0.648081570863724
12 -1.27211904525757
};
\addplot [semithick, darkslategray66]
table {%
12 -0.153626337647438
12 0.579345226287842
};
\addplot [semithick, darkslategray66]
table {%
11.8 -1.27211904525757
12.2 -1.27211904525757
};
\addplot [semithick, darkslategray66]
table {%
11.8 0.579345226287842
12.2 0.579345226287842
};
\addplot [black, mark=diamond*, mark size=2.5, mark options={solid,fill=darkslategray66}, only marks]
table {%
12 -1.99372124671936
12 -1.56695556640625
12 0.628438889980316
};
\addplot [semithick, darkslategray66]
table {%
13 -0.32913251966238
13 -1.18763327598572
};
\addplot [semithick, darkslategray66]
table {%
13 0.278933241963387
13 1.05335700511932
};
\addplot [semithick, darkslategray66]
table {%
12.8 -1.18763327598572
13.2 -1.18763327598572
};
\addplot [semithick, darkslategray66]
table {%
12.8 1.05335700511932
13.2 1.05335700511932
};
\addplot [black, mark=diamond*, mark size=2.5, mark options={solid,fill=darkslategray66}, only marks]
table {%
13 1.32789885997772
13 1.334233045578
13 1.20815587043762
13 1.61900997161865
};
\addplot [semithick, darkslategray66]
table {%
14 -0.381361827254295
14 -1.10113298892975
};
\addplot [semithick, darkslategray66]
table {%
14 0.493440732359886
14 1.62322819232941
};
\addplot [semithick, darkslategray66]
table {%
13.8 -1.10113298892975
14.2 -1.10113298892975
};
\addplot [semithick, darkslategray66]
table {%
13.8 1.62322819232941
14.2 1.62322819232941
};
\addplot [black, mark=diamond*, mark size=2.5, mark options={solid,fill=darkslategray66}, only marks]
table {%
14 2.50751924514771
};
\addplot [semithick, darkslategray66]
table {%
15 0.140507146716118
15 -0.993862211704254
};
\addplot [semithick, darkslategray66]
table {%
15 0.910328894853592
15 1.58431029319763
};
\addplot [semithick, darkslategray66]
table {%
14.8 -0.993862211704254
15.2 -0.993862211704254
};
\addplot [semithick, darkslategray66]
table {%
14.8 1.58431029319763
15.2 1.58431029319763
};
\addplot [black, mark=diamond*, mark size=2.5, mark options={solid,fill=darkslategray66}, only marks]
table {%
15 -1.27604985237122
15 -1.03265941143036
};
\addplot [semithick, darkslategray66]
table {%
-0.4 0.447769686579704
0.4 0.447769686579704
};
\addplot [semithick, darkslategray66]
table {%
0.6 -0.102737247943878
1.4 -0.102737247943878
};
\addplot [semithick, darkslategray66]
table {%
1.6 -1.45445883274078
2.4 -1.45445883274078
};
\addplot [semithick, darkslategray66]
table {%
2.6 1.15492856502533
3.4 1.15492856502533
};
\addplot [semithick, darkslategray66]
table {%
3.6 0.665297240018845
4.4 0.665297240018845
};
\addplot [semithick, darkslategray66]
table {%
4.6 -0.113554038107395
5.4 -0.113554038107395
};
\addplot [semithick, darkslategray66]
table {%
5.6 -0.983862161636353
6.4 -0.983862161636353
};
\addplot [semithick, darkslategray66]
table {%
6.6 0.587215572595596
7.4 0.587215572595596
};
\addplot [semithick, darkslategray66]
table {%
7.6 -0.608260542154312
8.4 -0.608260542154312
};
\addplot [semithick, darkslategray66]
table {%
8.6 -0.461860150098801
9.4 -0.461860150098801
};
\addplot [semithick, darkslategray66]
table {%
9.6 0.672729104757309
10.4 0.672729104757309
};
\addplot [semithick, darkslategray66]
table {%
10.6 0.632746398448944
11.4 0.632746398448944
};
\addplot [semithick, darkslategray66]
table {%
11.6 -0.408772453665733
12.4 -0.408772453665733
};
\addplot [semithick, darkslategray66]
table {%
12.6 -0.0540334284305573
13.4 -0.0540334284305573
};
\addplot [semithick, darkslategray66]
table {%
13.6 0.00915137678384781
14.4 0.00915137678384781
};
\addplot [semithick, darkslategray66]
table {%
14.6 0.489642545580864
15.4 0.489642545580864
};

\nextgroupplot[
height=\figheight,
tick align=outside,
tick pos=left,
title={Feature values (Classical reference)},
width=\figwidth,
x grid style={darkgray176},
xlabel={Latent feature},
xmin=-0.5, xmax=15.5,
xtick style={color=black},
y grid style={darkgray176},
ymajorgrids,
ymin=-30, ymax=35.494696521759,
ytick style={color=black},
ytick={-30, -20, -10, 0, 10, 20, 30}
]
\path [draw=darkslategray66, fill=steelblue46125174, semithick]
(axis cs:-0.4,-4.68981564044952)
--(axis cs:0.4,-4.68981564044952)
--(axis cs:0.4,-0.567691758275032)
--(axis cs:-0.4,-0.567691758275032)
--(axis cs:-0.4,-4.68981564044952)
--cycle;
\path [draw=darkslategray66, fill=dimgray10784138, semithick]
(axis cs:0.6,1.88290193676949)
--(axis cs:1.4,1.88290193676949)
--(axis cs:1.4,9.42720437049866)
--(axis cs:0.6,9.42720437049866)
--(axis cs:0.6,1.88290193676949)
--cycle;
\path [draw=darkslategray66, fill=salmon234109113, semithick]
(axis cs:1.6,-0.330201081931591)
--(axis cs:2.4,-0.330201081931591)
--(axis cs:2.4,6.58560192584991)
--(axis cs:1.6,6.58560192584991)
--(axis cs:1.6,-0.330201081931591)
--cycle;
\path [draw=darkslategray66, fill=gray, semithick]
(axis cs:2.6,-1.46035924553871)
--(axis cs:3.4,-1.46035924553871)
--(axis cs:3.4,0.899411112070084)
--(axis cs:2.6,0.899411112070084)
--(axis cs:2.6,-1.46035924553871)
--cycle;
\path [draw=darkslategray66, fill=steelblue46125174, semithick]
(axis cs:3.6,-3.56287920475006)
--(axis cs:4.4,-3.56287920475006)
--(axis cs:4.4,0.109577372670174)
--(axis cs:3.6,0.109577372670174)
--(axis cs:3.6,-3.56287920475006)
--cycle;
\path [draw=darkslategray66, fill=dimgray10784138, semithick]
(axis cs:4.6,0.384840562939644)
--(axis cs:5.4,0.384840562939644)
--(axis cs:5.4,7.26197457313538)
--(axis cs:4.6,7.26197457313538)
--(axis cs:4.6,0.384840562939644)
--cycle;
\path [draw=darkslategray66, fill=salmon234109113, semithick]
(axis cs:5.6,0.162284944206476)
--(axis cs:6.4,0.162284944206476)
--(axis cs:6.4,3.91322857141495)
--(axis cs:5.6,3.91322857141495)
--(axis cs:5.6,0.162284944206476)
--cycle;
\path [draw=darkslategray66, fill=gray, semithick]
(axis cs:6.6,-0.291819274425507)
--(axis cs:7.4,-0.291819274425507)
--(axis cs:7.4,2.71918946504593)
--(axis cs:6.6,2.71918946504593)
--(axis cs:6.6,-0.291819274425507)
--cycle;
\path [draw=darkslategray66, fill=steelblue46125174, semithick]
(axis cs:7.6,1.473741710186)
--(axis cs:8.4,1.473741710186)
--(axis cs:8.4,11.3615193367004)
--(axis cs:7.6,11.3615193367004)
--(axis cs:7.6,1.473741710186)
--cycle;
\path [draw=darkslategray66, fill=dimgray10784138, semithick]
(axis cs:8.6,-0.126780025660992)
--(axis cs:9.4,-0.126780025660992)
--(axis cs:9.4,4.2965487241745)
--(axis cs:8.6,4.2965487241745)
--(axis cs:8.6,-0.126780025660992)
--cycle;
\path [draw=darkslategray66, fill=salmon234109113, semithick]
(axis cs:9.6,-3.87415164709091)
--(axis cs:10.4,-3.87415164709091)
--(axis cs:10.4,0.346557542681694)
--(axis cs:9.6,0.346557542681694)
--(axis cs:9.6,-3.87415164709091)
--cycle;
\path [draw=darkslategray66, fill=gray, semithick]
(axis cs:10.6,0.20943459123373)
--(axis cs:11.4,0.20943459123373)
--(axis cs:11.4,4.76168715953827)
--(axis cs:10.6,4.76168715953827)
--(axis cs:10.6,0.20943459123373)
--cycle;
\path [draw=darkslategray66, fill=steelblue46125174, semithick]
(axis cs:11.6,-1.49471485614777)
--(axis cs:12.4,-1.49471485614777)
--(axis cs:12.4,7.87479615211487)
--(axis cs:11.6,7.87479615211487)
--(axis cs:11.6,-1.49471485614777)
--cycle;
\path [draw=darkslategray66, fill=dimgray10784138, semithick]
(axis cs:12.6,-3.17478978633881)
--(axis cs:13.4,-3.17478978633881)
--(axis cs:13.4,2.17071968317032)
--(axis cs:12.6,2.17071968317032)
--(axis cs:12.6,-3.17478978633881)
--cycle;
\path [draw=darkslategray66, fill=salmon234109113, semithick]
(axis cs:13.6,-3.0893212556839)
--(axis cs:14.4,-3.0893212556839)
--(axis cs:14.4,0.488471113145351)
--(axis cs:13.6,0.488471113145351)
--(axis cs:13.6,-3.0893212556839)
--cycle;
\path [draw=darkslategray66, fill=gray, semithick]
(axis cs:14.6,1.45430460572243)
--(axis cs:15.4,1.45430460572243)
--(axis cs:15.4,14.3129975795746)
--(axis cs:14.6,14.3129975795746)
--(axis cs:14.6,1.45430460572243)
--cycle;
\addplot [semithick, darkslategray66]
table {%
0 -4.68981564044952
0 -10.4607210159302
};
\addplot [semithick, darkslategray66]
table {%
0 -0.567691758275032
0 4.11832427978516
};
\addplot [semithick, darkslategray66]
table {%
-0.2 -10.4607210159302
0.2 -10.4607210159302
};
\addplot [semithick, darkslategray66]
table {%
-0.2 4.11832427978516
0.2 4.11832427978516
};
\addplot [black, mark=diamond*, mark size=2.5, mark options={solid,fill=darkslategray66}, only marks]
table {%
0 -12.6715383529663
0 -11.4846506118774
0 -11.9429311752319
0 -12.7819547653198
0 -11.7547035217285
};
\addplot [semithick, darkslategray66]
table {%
1 1.88290193676949
1 -1.3041969537735
};
\addplot [semithick, darkslategray66]
table {%
1 9.42720437049866
1 20.303092956543
};
\addplot [semithick, darkslategray66]
table {%
0.8 -1.3041969537735
1.2 -1.3041969537735
};
\addplot [semithick, darkslategray66]
table {%
0.8 20.303092956543
1.2 20.303092956543
};
\addplot [black, mark=diamond*, mark size=2.5, mark options={solid,fill=darkslategray66}, only marks]
table {%
1 22.1766948699951
1 20.8420295715332
1 30.4069862365723
};
\addplot [semithick, darkslategray66]
table {%
2 -0.330201081931591
2 -6.09844636917114
};
\addplot [semithick, darkslategray66]
table {%
2 6.58560192584991
2 16.7342796325684
};
\addplot [semithick, darkslategray66]
table {%
1.8 -6.09844636917114
2.2 -6.09844636917114
};
\addplot [semithick, darkslategray66]
table {%
1.8 16.7342796325684
2.2 16.7342796325684
};
\addplot [black, mark=diamond*, mark size=2.5, mark options={solid,fill=darkslategray66}, only marks]
table {%
2 18.4046993255615
2 17.6916275024414
2 20.2020702362061
2 18.137674331665
2 17.0646495819092
2 21.9984493255615
2 18.9583110809326
2 16.9619636535645
2 17.9605464935303
2 17.4217548370361
};
\addplot [semithick, darkslategray66]
table {%
3 -1.46035924553871
3 -4.97445344924927
};
\addplot [semithick, darkslategray66]
table {%
3 0.899411112070084
3 4.42588806152344
};
\addplot [semithick, darkslategray66]
table {%
2.8 -4.97445344924927
3.2 -4.97445344924927
};
\addplot [semithick, darkslategray66]
table {%
2.8 4.42588806152344
3.2 4.42588806152344
};
\addplot [black, mark=diamond*, mark size=2.5, mark options={solid,fill=darkslategray66}, only marks]
table {%
3 -7.58057069778442
3 -5.72184991836548
3 -5.7859845161438
3 -6.44965982437134
3 -8.56545543670654
3 -16.439754486084
3 -5.7689528465271
3 -9.09566593170166
3 -8.04465866088867
3 -9.04939460754395
3 -5.59884262084961
3 -12.4250621795654
3 4.68990230560303
3 4.62656497955322
3 4.86750602722168
3 4.66865396499634
3 5.51656246185303
3 4.61303424835205
3 4.91215515136719
3 5.13600397109985
3 6.03396654129028
3 4.47622871398926
3 5.28795194625854
};
\addplot [semithick, darkslategray66]
table {%
4 -3.56287920475006
4 -8.85488700866699
};
\addplot [semithick, darkslategray66]
table {%
4 0.109577372670174
4 5.52487516403198
};
\addplot [semithick, darkslategray66]
table {%
3.8 -8.85488700866699
4.2 -8.85488700866699
};
\addplot [semithick, darkslategray66]
table {%
3.8 5.52487516403198
4.2 5.52487516403198
};
\addplot [black, mark=diamond*, mark size=2.5, mark options={solid,fill=darkslategray66}, only marks]
table {%
4 -9.59709930419922
4 -9.53853893280029
4 -14.5464010238647
4 -12.0561571121216
4 -14.9449634552002
4 -14.6369209289551
4 -13.8748092651367
4 -10.1713333129883
4 -11.888331413269
4 5.64977121353149
4 7.17346668243408
4 6.52297687530518
4 5.87664842605591
4 5.78159093856812
};
\addplot [semithick, darkslategray66]
table {%
5 0.384840562939644
5 -4.74627113342285
};
\addplot [semithick, darkslategray66]
table {%
5 7.26197457313538
5 15.9808807373047
};
\addplot [semithick, darkslategray66]
table {%
4.8 -4.74627113342285
5.2 -4.74627113342285
};
\addplot [semithick, darkslategray66]
table {%
4.8 15.9808807373047
5.2 15.9808807373047
};
\addplot [black, mark=diamond*, mark size=2.5, mark options={solid,fill=darkslategray66}, only marks]
table {%
5 -10.4090538024902
};
\addplot [semithick, darkslategray66]
table {%
6 0.162284944206476
6 -4.65447664260864
};
\addplot [semithick, darkslategray66]
table {%
6 3.91322857141495
6 9.16385078430176
};
\addplot [semithick, darkslategray66]
table {%
5.8 -4.65447664260864
6.2 -4.65447664260864
};
\addplot [semithick, darkslategray66]
table {%
5.8 9.16385078430176
6.2 9.16385078430176
};
\addplot [black, mark=diamond*, mark size=2.5, mark options={solid,fill=darkslategray66}, only marks]
table {%
6 10.7690439224243
6 12.6770429611206
6 10.8189430236816
6 12.0011987686157
6 9.8471794128418
6 11.7536811828613
6 11.1047811508179
6 10.2408599853516
6 12.4435815811157
6 10.6182613372803
6 12.070538520813
6 10.9169521331787
};
\addplot [semithick, darkslategray66]
table {%
7 -0.291819274425507
7 -4.20405006408691
};
\addplot [semithick, darkslategray66]
table {%
7 2.71918946504593
7 6.71830320358276
};
\addplot [semithick, darkslategray66]
table {%
6.8 -4.20405006408691
7.2 -4.20405006408691
};
\addplot [semithick, darkslategray66]
table {%
6.8 6.71830320358276
7.2 6.71830320358276
};
\addplot [black, mark=diamond*, mark size=2.5, mark options={solid,fill=darkslategray66}, only marks]
table {%
7 -7.07742023468018
7 -7.77351760864258
7 -6.59443378448486
7 -5.58308410644531
7 -5.07581615447998
7 7.92848062515259
7 7.23608016967773
7 10.5644245147705
7 7.95626258850098
7 8.25555419921875
7 10.5327005386353
7 8.52611923217773
7 9.65668869018555
7 8.59225368499756
7 10.3181467056274
7 13.646782875061
7 7.59269618988037
7 7.34428071975708
7 7.41745948791504
7 8.01222324371338
};
\addplot [semithick, darkslategray66]
table {%
8 1.473741710186
8 -2.80567216873169
};
\addplot [semithick, darkslategray66]
table {%
8 11.3615193367004
8 21.8806133270264
};
\addplot [semithick, darkslategray66]
table {%
7.8 -2.80567216873169
8.2 -2.80567216873169
};
\addplot [semithick, darkslategray66]
table {%
7.8 21.8806133270264
8.2 21.8806133270264
};
\addplot [black, mark=diamond*, mark size=2.5, mark options={solid,fill=darkslategray66}, only marks]
table {%
8 26.723690032959
};
\addplot [semithick, darkslategray66]
table {%
9 -0.126780025660992
9 -5.89048433303833
};
\addplot [semithick, darkslategray66]
table {%
9 4.2965487241745
9 10.2620801925659
};
\addplot [semithick, darkslategray66]
table {%
8.8 -5.89048433303833
9.2 -5.89048433303833
};
\addplot [semithick, darkslategray66]
table {%
8.8 10.2620801925659
9.2 10.2620801925659
};
\addplot [black, mark=diamond*, mark size=2.5, mark options={solid,fill=darkslategray66}, only marks]
table {%
9 -6.9507360458374
9 11.2776670455933
9 19.8123111724854
9 13.4812965393066
};
\addplot [semithick, darkslategray66]
table {%
10 -3.87415164709091
10 -9.35979461669922
};
\addplot [semithick, darkslategray66]
table {%
10 0.346557542681694
10 6.63579416275024
};
\addplot [semithick, darkslategray66]
table {%
9.8 -9.35979461669922
10.2 -9.35979461669922
};
\addplot [semithick, darkslategray66]
table {%
9.8 6.63579416275024
10.2 6.63579416275024
};
\addplot [black, mark=diamond*, mark size=2.5, mark options={solid,fill=darkslategray66}, only marks]
table {%
10 -20.6017875671387
10 -11.4504842758179
10 -12.4050102233887
10 6.70768928527832
10 7.22499370574951
10 11.9660959243774
10 12.9837217330933
10 6.96402835845947
10 6.67967700958252
};
\addplot [semithick, darkslategray66]
table {%
11 0.20943459123373
11 -6.23526859283447
};
\addplot [semithick, darkslategray66]
table {%
11 4.76168715953827
11 10.5354690551758
};
\addplot [semithick, darkslategray66]
table {%
10.8 -6.23526859283447
11.2 -6.23526859283447
};
\addplot [semithick, darkslategray66]
table {%
10.8 10.5354690551758
11.2 10.5354690551758
};
\addplot [black, mark=diamond*, mark size=2.5, mark options={solid,fill=darkslategray66}, only marks]
table {%
11 -6.69056510925293
11 -7.70812320709229
11 -7.69982814788818
11 -8.7929801940918
11 12.6100645065308
11 14.1396570205688
11 15.4927368164062
11 13.6752634048462
11 11.9033174514771
};
\addplot [semithick, darkslategray66]
table {%
12 -1.49471485614777
12 -14.7266693115234
};
\addplot [semithick, darkslategray66]
table {%
12 7.87479615211487
12 21.2272319793701
};
\addplot [semithick, darkslategray66]
table {%
11.8 -14.7266693115234
12.2 -14.7266693115234
};
\addplot [semithick, darkslategray66]
table {%
11.8 21.2272319793701
12.2 21.2272319793701
};
\addplot [black, mark=diamond*, mark size=2.5, mark options={solid,fill=darkslategray66}, only marks]
table {%
12 -18.2562789916992
12 -16.1179580688477
12 -16.0712852478027
12 -15.7891550064087
12 -20.8756275177002
12 -20.4537372589111
12 -19.2042045593262
12 -22.9950847625732
12 -19.35085105896
12 24.0995388031006
};
\addplot [semithick, darkslategray66]
table {%
13 -3.17478978633881
13 -10.2206134796143
};
\addplot [semithick, darkslategray66]
table {%
13 2.17071968317032
13 10.1362476348877
};
\addplot [semithick, darkslategray66]
table {%
12.8 -10.2206134796143
13.2 -10.2206134796143
};
\addplot [semithick, darkslategray66]
table {%
12.8 10.1362476348877
13.2 10.1362476348877
};
\addplot [black, mark=diamond*, mark size=2.5, mark options={solid,fill=darkslategray66}, only marks]
table {%
13 -11.5269775390625
13 12.9831504821777
13 14.1575212478638
13 18.6070289611816
13 16.168140411377
13 10.8401393890381
13 13.7055892944336
13 10.6109104156494
13 18.0410823822021
13 13.9125604629517
13 15.3944787979126
13 15.2271776199341
13 18.0984764099121
13 21.2560157775879
13 17.0722045898438
13 16.0137405395508
13 19.5536441802979
13 14.2082672119141
13 20.8031139373779
13 18.509090423584
13 11.8685932159424
13 15.4292392730713
13 11.8231973648071
13 11.361780166626
13 24.6079387664795
13 22.0885734558105
};
\addplot [semithick, darkslategray66]
table {%
14 -3.0893212556839
14 -8.30835151672363
};
\addplot [semithick, darkslategray66]
table {%
14 0.488471113145351
14 5.66273593902588
};
\addplot [semithick, darkslategray66]
table {%
13.8 -8.30835151672363
14.2 -8.30835151672363
};
\addplot [semithick, darkslategray66]
table {%
13.8 5.66273593902588
14.2 5.66273593902588
};
\addplot [black, mark=diamond*, mark size=2.5, mark options={solid,fill=darkslategray66}, only marks]
table {%
14 -8.83729076385498
14 -9.11678504943848
14 -9.48213386535645
14 -13.9140434265137
14 -10.9211521148682
14 7.38895416259766
14 6.88835334777832
14 8.66575622558594
};
\addplot [semithick, darkslategray66]
table {%
15 1.45430460572243
15 -4.17432498931885
};
\addplot [semithick, darkslategray66]
table {%
15 14.3129975795746
15 32.7094688415527
};
\addplot [semithick, darkslategray66]
table {%
14.8 -4.17432498931885
15.2 -4.17432498931885
};
\addplot [semithick, darkslategray66]
table {%
14.8 32.7094688415527
15.2 32.7094688415527
};
\addplot [semithick, darkslategray66]
table {%
-0.4 -1.8059606552124
0.4 -1.8059606552124
};
\addplot [semithick, darkslategray66]
table {%
0.6 6.24640965461731
1.4 6.24640965461731
};
\addplot [semithick, darkslategray66]
table {%
1.6 1.49505603313446
2.4 1.49505603313446
};
\addplot [semithick, darkslategray66]
table {%
2.6 -0.197417661547661
3.4 -0.197417661547661
};
\addplot [semithick, darkslategray66]
table {%
3.6 -0.850412666797638
4.4 -0.850412666797638
};
\addplot [semithick, darkslategray66]
table {%
4.6 3.76122856140137
5.4 3.76122856140137
};
\addplot [semithick, darkslategray66]
table {%
5.6 1.33581018447876
6.4 1.33581018447876
};
\addplot [semithick, darkslategray66]
table {%
6.6 0.520870506763458
7.4 0.520870506763458
};
\addplot [semithick, darkslategray66]
table {%
7.6 6.54993844032288
8.4 6.54993844032288
};
\addplot [semithick, darkslategray66]
table {%
8.6 0.823802202939987
9.4 0.823802202939987
};
\addplot [semithick, darkslategray66]
table {%
9.6 -0.689203888177872
10.4 -0.689203888177872
};
\addplot [semithick, darkslategray66]
table {%
10.6 1.68329393863678
11.4 1.68329393863678
};
\addplot [semithick, darkslategray66]
table {%
11.6 0.661084711551666
12.4 0.661084711551666
};
\addplot [semithick, darkslategray66]
table {%
12.6 -0.333040162920952
13.4 -0.333040162920952
};
\addplot [semithick, darkslategray66]
table {%
13.6 -0.55023193359375
14.4 -0.55023193359375
};
\addplot [semithick, darkslategray66]
table {%
14.6 7.01156854629517
15.4 7.01156854629517
};
\end{groupplot}

\end{tikzpicture}